\documentclass[iop]{emulateapj}

\usepackage{amsmath, amsthm, amssymb, slashed}
\usepackage{color}
\usepackage[toc]{appendix}
\newcommand{\dst}{\displaystyle}
\newcommand{\ada}{a+\delta a}
\newcommand{\zs}{z_{\rm s}}

\newcommand{\laka}{(\lambda, \kappa)}
\newcommand{\rp}{{\rho}^{\prime}}
\newcommand{\dd}{{\rm d}}
\newcommand{\ii}{{\rm i}}

\shorttitle{Local Interstellar Magnetic Field}
\shortauthors{R\"oken et al.}

\begin{document}


\title{An exact analytical solution for the interstellar magnetic field\\
       in the vicinity of the heliosphere}


\author{Christian R\"oken} 
\affil{Universit\"at Regensburg, Fakult\"at f\"ur Mathematik, 
       Regensburg, Germany}  
\email{christian.roeken@mathematik.uni-regensburg.de}

\author{Jens Kleimann and Horst Fichtner}
\affil{Ruhr-Universit\"at Bochum, Fakult\"at f\"ur Physik und Astronomie,
       Institut f\"ur Theoretische Physik IV, Bochum, Germany}
\email{jk@tp4.rub.de, hf@tp4.rub.de}



\begin{abstract}
An analytical representation of the interstellar magnetic field in the vicinity
of the heliosphere is derived. The three-dimensional field structure close to
the heliopause is calculated as a solution of the induction equation under the
assumption that it is frozen into a prescribed plasma flow resembling the
characteristic interaction of the solar wind with the local interstellar 
medium. The usefulness of this analytical solution as an approximation to 
self-consistent magnetic field configurations obtained numerically from the 
full MHD equations is illustrated by quantitative comparisons. 
\end{abstract}





\section{Introduction and Motivation}

With the likely entry of the {\em Voyager} spacecraft into interstellar space 
\citep{Gurnett-etal-2013}, with the recent measurements of the
{\em Interstellar Boundary Explorer (IBEX)} that constrain the physical
properties of the local 
interstellar medium (LISM, see the reviews by \citet{McComas-etal-2012} and  
\citet{McComas-etal-2014}), and with the notion that the so-called heliotail 
may be of significance for anisotropies in the flux of galactic cosmic rays 
\citep{Amenomori-etal-2010,Desiati-Lazarian-2013,Schwadron-etal-2014}, the
nature of the local interstellar magnetic field (ISMF) has recently received
increased attention. Prior to these new measurements, which are related to
regions outside but close to the heliosphere, the ISMF has either been
investigated in a rather astrophysical context, i.e., as the local 
representation of the general galactic magnetic field 
\citep[e.g.][]{Amenomori-etal-2006,Frisch-2007} or as an outer `boundary
condition' for models with which an asymmetry in the large-scale structure
of the heliosphere was studied \citep[e.g.][]{Izmodenov-etal-2005b,
Opher-etal-2007, Ratkiewicz-Grygorczuk-2008, Pogorelov-etal-2009}.

Particularly for the latter application, sophisticated three-dimensional
magnetohydrodynamics (MHD)
\citep[e.g.][]{BenJaffel-etal-2013}, multi-fluid plasma-neutral 
\citep[e.g.][]{Opher-Drake-2013, Borovikov-Pogorelov-2014}, and MHD-kinetic
models \citep{Heerikhuisen-etal-2008,Zank-etal-2013} have been developed and 
result in a `realistic' three-dimensional structuring of the ISMF in the
vicinity of the heliosphere as a consequence of a `draping' of field lines
over the heliopause, as already described conceptually by
\citet{Belcher-etal-1993}. While such fully numerical computations of the
local ISMF are required for detailed comparisons of model simulations with 
measurements, they are not suitable for all purposes as is, e.g. discussed 
in \citet{Mitchell-etal-2008}. An example is the recent work by 
\citet{Schwadron-etal-2014}, where an approximation of the local ISMF has
been used in order to compute trajectories of galactic cosmic rays.  

Approximations of the local ISMF that is perturbed by the presence of the
heliosphere are as old as the concept of the heliosphere itself. Already 
\citet{Parker-1961} derived the first non-trivial, non-flow-parallel ISMF
configuration by neglecting the interstellar flow field. Similar approaches
have been used by various authors over the years and are still in use, see for
example the application of the line dipole method by \citet{Whang-2010} or
the magnetic potential representation employed by \citet{Schwadron-etal-2014}.
A common feature  of these approximations is the neglect of an explicitly
treated plasma flow and  the prescription of the heliopause surface on purely
magnetic (line of dipoles) or geometric (spherically capped cylinder) grounds.
A first improvement was presented by \citet{Mitchell-etal-2008} who, by
exploiting  the frozen-in condition, numerically computed the ISMF for a
prescribed plasma  flow that was taken from a numerical simulation by
\citet{Zank-etal-1996b}. 

To the best of our knowledge, a fully analytical calculation of the ISMF
frozen into a plasma flow resulting from the interaction of the interstellar
flow with the solar wind has not been treated in the literature. With the
present
paper, we fill this gap with analytically calculating the
three-dimensional ISMF structure in the vicinity of the heliosphere by
assuming a plasma flow field considered to be typical for the
heliosphere--LISM interaction. The unperturbed frozen-in ISMF at large
distances is allowed to have an arbitrary inclination relative to the 
upwind--downwind axis of the heliosphere.

This paper is structured as follows: In Section~2, the plasma flow field being
characteristic for the interaction of the solar wind with the LISM is defined
and  in Section~3, the resulting frozen-in ISMF is calculated.
In Section~4, a comparison of this analytical solution with results from
numerical simulations is presented and critically discussed, and a summary
of the main results is given in the concluding Section~5.

\section{The Interaction Scenario Between the
  Heliosphere and the LISM}

The outer boundary of the heliosphere, the heliopause, is determined as the
separatrix between the solar wind plasma and the interstellar plasma flow.
In an approximation that is very useful for many purposes, the flow
velocity ${\mathbf u}$ in the vicinity of the heliopause can be considered as 
incompressible ($ \boldsymbol{\nabla}\cdot{\mathbf u}=0$) and irrotational
($ \boldsymbol{\nabla}\times{\mathbf u}={\mathbf 0}$), resulting
from the superposition of a radial flow emanating from a stationary
point-like source and a homogeneous flow from infinity. This can be
formulated via a scalar velocity potential
\begin{equation}
  \Phi({\mathbf r}) = u_0 \left( z + \frac{q}{r} \right)
\end{equation}
at position ${\mathbf r} = \rho \, {\mathbf e}_{\rho} + z {\mathbf e}_z$
(where $\rho$ and $z$ denote cylindrical coordinates with corresponding
orthogonal unit vectors ${\mathbf e}_{\rho,z}$, and
$r := ||\mathbf r|| = \sqrt{\rho^2+z^2}$), from which the velocity
field
\begin{equation} \label{eq:u-flow}
  {\bf u}({\mathbf r}) = -  \boldsymbol{\nabla} \Phi({\bf r}) = u_0 \left[
    \frac{q \, \rho}{r^3} {\mathbf e}_{\rho}
    + \left( \frac{q \, z}{r^3} -1 \right)
    {\mathbf e}_z \right]
\end{equation}
is then derived. 
This Rankine-type flow was first proposed as a heliospheric flow model
by \citet{Parker-1961}.
The two constants $u_0$ and $q$ represent the speed of the
homogeneous interstellar flow (incident from the positive $z$ direction) and
the relative strength of the point-like solar wind source, respectively.
It should be noted that by normalizing all lengths to
\mbox{$L_{\rm s}:=\sqrt{q}$},
the $q$ dependence can be removed completely, or in other words, a change in
source
strength $q \rightarrow q^{\prime}$ will cause all lengths to expand by a
factor $q^{\prime}/q$ while conserving the overall shape of the flow.
However, for the sake of dimensional clarity, we chose to retain this
dependence throughout the calculations.

\begin{figure}[ht!]
  \epsscale{1.15}
  \plotone{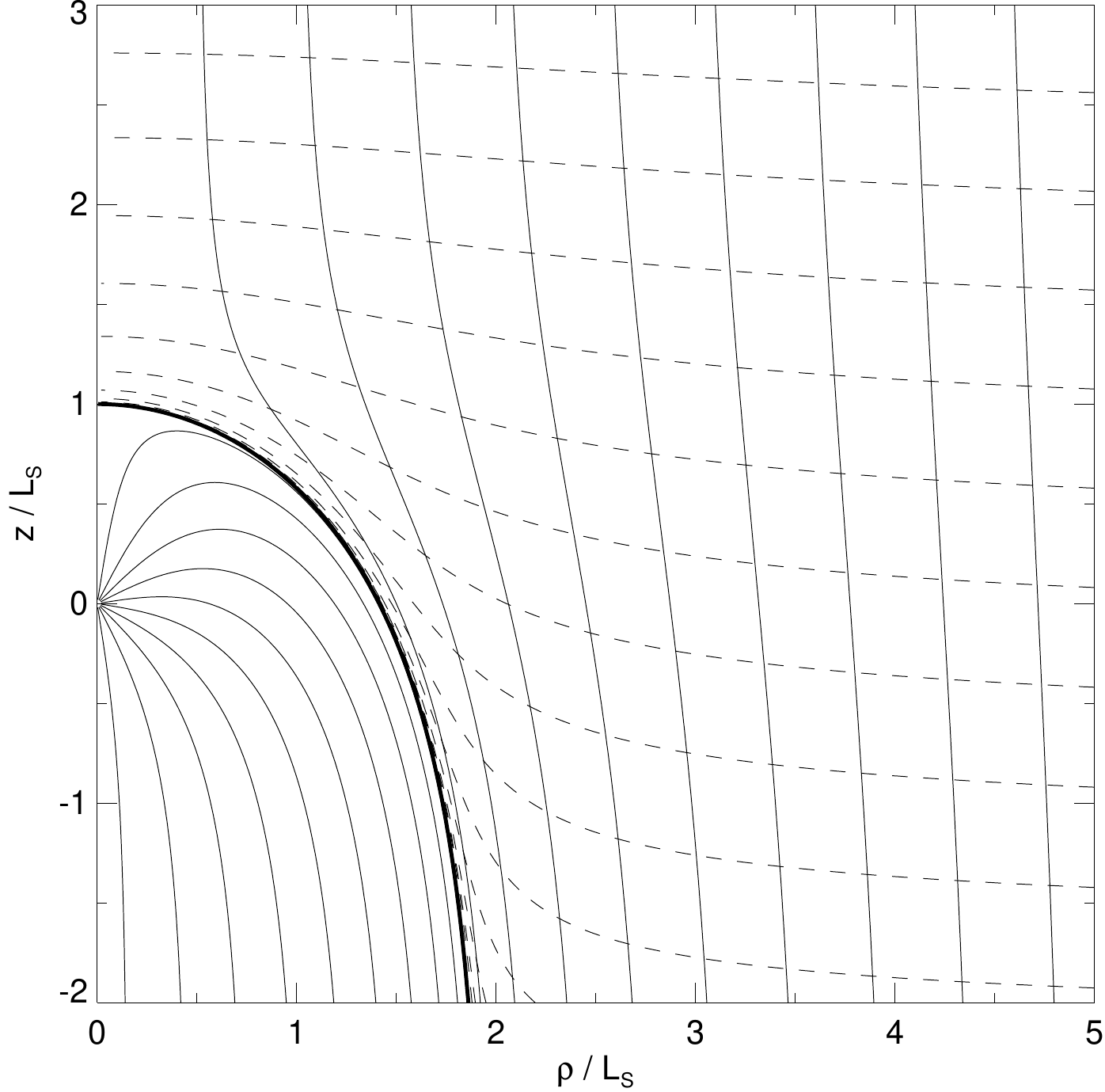}
  \caption{ \label{fig:flowsketch}
    The streamlines of the flow field (\ref{eq:u-flow}) (solid), plotted as
    lines of constant $a$ using Eq.~(\ref{eq:za}) in the rest frame of the
    Sun located at the origin. The dashed lines are isochrones, computed
    by substituting $\varphi=0$ and $B_{y 0} = 0 = B_{z 0}$ into
    Eqs.~(\ref{finalr}) and (\ref{finalz}), and then numerically integrating
    $\dd z / \dd \rho = B_z/B_{\rho}$ from starting points at $\rho_0=8$,
    $z_0 \in \{-4, -3.5, \ldots, 4 \}$ towards smaller $\rho$.
    The heliopause is visible as the thick, solid line through the stagnation
    point. Coordinates $\rho$ and $z$ are normalized to the standoff distance
    $L_{\rm s} = \sqrt{q}$.
  }
\end{figure}

Streamlines for this flow field are computed as solutions to the equation
\begin{equation} \label{eq:def-flowlines}
  \frac{\dd z}{\dd \rho} = \frac{u_z}{u_{\rho}} \ ,
\end{equation}
which read
\begin{equation}
  \label{eq:za}
  z_a(\rho) = \frac{ \left( 2q +a^2-\rho^2 \right) \rho}
  {\sqrt{4 q^2 - \left( 2q +a^2-\rho^2 \right)^2}} \ .
\end{equation}
Here, the parameter $a$ denotes the (asymptotic) distance of a streamline to
the axis $\rho = 0$ for $z\to\infty$. Evidently, $a$ can be used to label streamlines,
which will be exploited in Section~\ref{sec:sol_jk}. For any streamline $a$,
$\rho$ varies monotonously from $a$ to $\sqrt{a^2+4q}$ (the latter value
being only assumed asymptotically in the limit $z\to -\infty$).
Selected streamlines 
are illustrated in Fig.~\ref{fig:flowsketch}, together
with isochrones (i.e., lines connecting flow elements which started at a
common point of time at infinite $z$).
In this parameterization, the heliopause, indicated by the thick solid line in
Fig.~\ref{fig:flowsketch}, corresponds to the particular streamline $a=0$,
while all solar wind streamlines (i.e., those internal to the heliopause)
have imaginary $a$ values, and are not addressed in this paper.

As the prescribed stationary flow field (\ref{eq:u-flow}) is
divergence-free, it
represents an incompressible interstellar and solar wind flow. This
condition is a good approximation for the subsonic solar wind in the inner
heliosheath between the termination shock and the heliopause. It should also
be a suitable approximation for the interstellar flow downstream of the
heliospheric bow shock. While even for a configuration without a bow shock
\citep{McComas-etal-2012b} the streamlines are not expected to be much
different, such shock is however likely to exist
\citep{BenJaffel-etal-2013, Scherer-Fichtner-2014}.
\section{The Analytical Solution}

In the following, we derive an exact solution for the ISMF that is treated to
be time-independent,
homogeneous at infinity, and frozen into the interstellar flow of
Eq.~(\ref{eq:u-flow}). The latter assumption limits  the validity of the
solution to regions where the dynamics of the plasma flow is not dominated by
the ISMF. While very close to the heliopause this limitation will be
violated, it is demonstrated in Section~4 that the solution is, nonetheless,
a valid and useful approximation to self-consistent field configurations
obtained numerically from the full set of MHD equations.     
\subsection{Boundary Conditions}
\label{sec:bconds}
An outer boundary condition is prescribed at infinity where the ISMF should
be homogeneous, i.e., in Cartesian coordinates  
${\bf B}_0 = B_{x0}{\bf e}_{x} + B_{y0}{\bf e}_{y}
+ B_{z0}{\bf e}_{z}$ with constants $B_{x0}$, $B_{y0}$, and $B_{z0}$ holds.

An inner boundary condition for the ISMF is given at the heliopause to which
it must be tangential. This is intrinsically fulfilled by the use of the
frozen-in condition.  
\subsection{Derivation from a Set of Basic Partial
  Differential Equations}
Starting with the frozen-in condition, the steady-state induction equation
reads
\begin{equation}\label{TIIMHDOL}
   \boldsymbol{\nabla} \times (\mathbf{u} \times \mathbf{B}) = \boldsymbol{0} \ .
\end{equation}
With the solenoidality constraint
\begin{equation}\label{eq:divB}
   \boldsymbol{\nabla} \cdot \mathbf{B} = 0
\end{equation}
and the incompressibility condition, Eq.~(\ref{TIIMHDOL}) simplifies to
\begin{equation}\label{eq:FL}
  (\mathbf{B} \cdot  \boldsymbol{\nabla}) \mathbf{u} =
  (\mathbf{u} \cdot  \boldsymbol{\nabla}) \mathbf{B} \ .
\end{equation}
Because the region exterior to the heliosphere is simply connected, the
Poincar\'e lemma states that the curl-free vector field
$\mathbf{u} \times \mathbf{B}$ can be represented by the gradient field
of a potential $\Psi$
\begin{equation}\label{POTREP}
  \mathbf{u} \times \mathbf{B} = \boldsymbol{\nabla} \Psi \ .
\end{equation}

By using this representation, the level of difficulty in finding the magnetic
field solution from Eq.~(\ref{eq:FL}) can be considerably reduced.
Hence, in order to find explicit analytical ISMF solutions, one has to solve
the set of coupled partial differential equations (PDEs) for Eq.~(\ref{eq:FL}) 
and Eq.~(\ref{POTREP}), which in cylindrical coordinates $(\rho,\varphi,z)$ 
read
\begin{align}
  & \left(\rho \partial_{\rho} + \left[z - \frac{r^3}{q} \right] \partial_z - 1\right) B_{\varphi} = 0 \label{DGL1} \\ 
  & r^2 \left(\rho \partial_{\rho} + \left[z - \frac{r^3}{q}\right] \partial_z\right) B_{\rho} = (z^2 - 2 \rho^2) B_{\rho} - 3 \rho z B_{z} \label{DGL2} \\ 
  & r^2 \left(\rho \partial_{\rho} + \left[z - \frac{r^3}{q}\right] \partial_z\right) B_{z} = (\rho^2 - 2 z^2) B_{z} - 3 \rho z B_{\rho} \label{DGL3} \\  
  % & \partial_{\rho} (\rho B_{\rho}) + \partial_{\varphi} B_{\varphi} + \rho \partial_z B_z = 0 \label{DGL4} \\  
  & \partial_{\rho} \Psi = - \frac{u_0 q}{r^3} \, \left[z - \frac{r^3}{q}\right] B_{\varphi} \label{DGL5} \\  
  & \partial_{z} \Psi = \frac{u_0 q \, \rho}{r^3} \, B_{\varphi} \label{DGL6} \\  
  & \partial_{\varphi} \Psi = \frac{u_0 q \, \rho}{r^3} \left(\left[z - \frac{r^3}{q}\right] B_{\rho} - \rho B_z\right). \label{DGL7} 
\end{align}
\subsubsection{Angular Component of the ISMF}
From the first PDE Eq.~(\ref{DGL1}), it can directly be seen that the
$\varphi$ component of the ISMF is already decoupled. Applying the spherical
coordinate transformation
$(\rho, z) \mapsto (r, \vartheta)$, $r \in \mathbb{R}_{> 0}$ and
$ \vartheta \in [- \pi/2, \pi/2]$, with
\begin{equation}\label{CT1}
  \rho = r \cos{(\vartheta)} \quad \textnormal{and} \quad
  z = r \sin{(\vartheta)}
\end{equation}
as well as
\begin{equation}\label{PD1}
  \begin{split}
    \partial_{\rho} &= \cos{(\vartheta)} \
    \partial_r - \frac{\sin{(\vartheta)}}{r} \ \partial_{\vartheta} \\
    \partial_z &= \sin{(\vartheta)} \
    \partial_r + \frac{\cos{(\vartheta)}}{r} \ \partial_{\vartheta} \ ,
  \end{split}
\end{equation}
yields
\begin{equation}
  \left( \left[ 1 - \frac{r^2 \sin{(\vartheta)}}{q} \right] \partial_r
  - \frac{r \cos{(\vartheta)}}{q} \ \partial_{\vartheta}
  - \frac{1}{r} \right) B_{\varphi} = 0
\end{equation}
as a spherical representation of Eq.~(\ref{DGL1}).
To eliminate the $1/r$ term, and since this PDE does not
depend on the angular variable $\varphi$, one can use the product ansatz
$B_{\varphi} = B_{\varphi}(r, \vartheta, \varphi)
= r \cos{(\vartheta)} \mathcal{D}(r, \vartheta) \mathcal{H}(\varphi)$ 
to obtain a PDE for the function $\mathcal{D}(r, \vartheta)$ 
\begin{equation}\label{Bphi}
  \biggl(\biggl[r \tan{(\vartheta)}
  - \frac{q}{r \cos{(\vartheta)}}\biggr] \partial_r
  + \partial_{\vartheta}\biggr) \mathcal{D} = 0 \ .
\end{equation}
This PDE can be solved by the method of characteristics as
follows. Given a parameterization $(r(u,v), \vartheta(u,v))$ and a solution
$\mathcal{D}$ such that 
\begin{equation}\label{PDED}
  \begin{split}
    \partial_u \mathcal{D} &= (\partial_r \mathcal{D}) \,
    \frac{\dd r}{\dd u}
    + (\partial_{\vartheta} \mathcal{D}) \,
    \frac{\dd \vartheta}{\dd u} \\
    &= \left[ r \tan{(\vartheta)} - \frac{q}{r \cos{(\vartheta)}} \right]
    \partial_r \mathcal{D} + \partial_{\vartheta} \mathcal{D} \ ,
  \end{split}
\end{equation}
one can re-write the PDE (\ref{Bphi}) in terms of a family of ordinary
differential equations (ODEs) by equating the coefficients
\begin{eqnarray}
  \frac{\dd r}{\dd u} &=& r \tan{(\vartheta)} - \frac{q}{r \cos{(\vartheta)}} \label{F1} \\ 
  \frac{\dd \vartheta}{\dd u} &=& 1 \quad \Rightarrow \quad
  \vartheta = u + \vartheta_0(v) \label{F2} \\
  \partial_u \mathcal{D} &=& 0 \quad \Rightarrow \quad
  \mathcal{D} = \mathcal{F}(v) \label{F3} \ .
\end{eqnarray}
Substituting Eq.~(\ref{F2}) into Eq.~(\ref{F1}) yields the ODE
\begin{equation}\label{ODERHO}
  \rho(u, v) \, \frac{\dd \rho(u,v)}{\dd u}
  + q \cos{(u + \vartheta_0(v))} = 0
\end{equation}
for $\rho(u, v) = r(u, v) \cos{(u + \vartheta_0(v))}$, which is solved
straightforwardly by integration. The solution of Eq.~(\ref{ODERHO}), in
implicit form, reads
\begin{equation}\label{rsol}
  \frac{\rho^{2}(u, v)}{2} + q \sin{(u + \vartheta_0(v))} = \omega_0(v) \ ,
\end{equation}
and the function $\mathcal{D}$ thus becomes 
\begin{equation}
  \mathcal{D} = \mathcal{F}(v) = 
  \mathcal{F} \circ \omega_0^{[-1]}\left(\frac{\rho^2}{2} +\frac{q z}{r}\right)
  = \mathcal{G}\biggl(\frac{\rho^2}{2} + \frac{q z}{r}\biggr),
\end{equation}
where $\omega_0^{[-1]}$ denotes the inverse of $\omega_0$, and
$\mathcal{G}$ is a $C^1$ function yet to be determined.
Then, one obtains for the $\varphi$ component of the magnetic field the
expression
\begin{equation}\label{AMFC}
  B_{\varphi}(\rho, \varphi, z) = \rho \, \mathcal{G}\biggl(\frac{\rho^2}{2}
  + \frac{q z}{r}\biggr) \, \mathcal{H}(\varphi) \ .
\end{equation}
Because of the assumed homogeneity of ${\bf B}$ at infinity, the boundary
conditions for the $\varphi$ component are given by 
\begin{eqnarray}
  \label{LIM2} \lim_{\rho \rightarrow \infty} \rho \, \mathcal{G} &=& 1 \\ 
  \label{LIM1} \lim_{z \rightarrow \infty} \mathcal{G} &=& \frac{1}{\rho}
\end{eqnarray}
\begin{equation}
  \label{HFCT}
  \mathcal{H} = - \sin{(\varphi)} B_{x 0} + \cos{(\varphi)} B_{y 0} \ ,
\end{equation}
where $B_{x 0}$ and $B_{y 0}$ denote the constant Cartesian magnetic field
components introduced in Section~\ref{sec:bconds}.
Due to the global continuity of the ISMF, and therefore of the function
$\mathcal{G}$, the limit in Eq.~(\ref{LIM1}) can be pulled into the argument
of $\mathcal{G}$. One can deduce that
\begin{equation}
  \mathcal{G} \left(\frac{\rho^2}{2} + q\right) = \frac{1}{\rho} \ .
\end{equation}
From this condition and the general form of the argument of $\mathcal{G}$,
the latter function can be uniquely determined to be 
\begin{equation} \label{GF}
  \mathcal{G} \left( \frac{\rho^2}{2} + \frac{q z}{r} \right)
  = \frac{1}{\sqrt{\rho^2 + 2 q \left(\dst \frac{z}{r} - 1\right)}} \ .
\end{equation}
The remaining limit (\ref{LIM2}) is also fulfilled by Eq.~(\ref{GF}).
The angular component of the magnetic field (\ref{AMFC}) is therefore
fixed by the homogeneity conditions at infinity, yielding 
\begin{equation}\label{AMFC2}
  B_{\varphi}(\rho, \varphi, z) = \frac{\rho \,
    \bigl(- \sin{(\varphi)} B_{x 0}
    + \cos{(\varphi)} B_{y 0}\bigr)}{\sqrt{\rho^2
      + 2 q \left( \dst \frac{z}{r} - 1 \right)}} \ .
\end{equation}

\subsubsection{Radial and Axial Components of the ISMF} 

Examining the coupled system of first-order PDEs given by Eqs.~(\ref{DGL2})
and (\ref{DGL3}), one can immediately see that they can be easily decoupled
via a simple algebraic manipulation for example of Eq.~(\ref{DGL3}) 
\begin{equation}\label{MFZ}
  B_{\rho} = - \frac{r^2 }{3 \rho z} \left( \rho \partial_{\rho}
    + \left[ z - \frac{r^3}{q} \right] \partial_z
    + \frac{2 z^2 - \rho^2}{r^2} \right) B_{z}
\end{equation}
and substitution into Eq.~(\ref{DGL2}), leading to a linear, homogeneous,
parabolic second-order PDE for the $B_z$ component 
\begin{eqnarray}\label{MFZ2}
  && \left( \rho^2 \partial_{\rho \rho}
  + \biggl[ z - \frac{r^3}{q} \right]^2 \partial_{z z}
  + 2 \rho \left[ z - \frac{r^3}{q} \right] \partial_{\rho z} \nonumber\\
  &&+ \rho \left[ 2 + \frac{r}{q z} (\rho^2 - z^2) \right] \partial_{\rho}
  \nonumber \\ 
  &&+ \left[ 2 z - \frac{2 r}{q} (2 \rho^2 + 3 z^2)
  - \frac{r^4}{q^2 z} (\rho^2 - 4 z^2) \right] \partial_z \nonumber\\
  &&- 2 - \frac{r}{q z} (\rho^2 + 2 z^2)\biggr) B_z = 0 \ .
\end{eqnarray}
The solution of this equation can be re-substituted into Eq.~(\ref{MFZ}) in
order to determine the remaining component $B_{\rho}$.
Solving this intricate second-order PDE directly can be avoided by first
inserting the $\varphi$ component of the ISMF (\ref{AMFC2}) into
Eqs.~(\ref{DGL5}) and (\ref{DGL6}), determining the function $\Psi$, and
substituting it into Eq.~(\ref{DGL7}) to explicitly relate $B_z$ to
$B_{\rho}$. Having the expression $B_z = B_z(B_{\rho})$, Eqs.~(\ref{DGL2}) and
(\ref{DGL3}) reduce to non-coupled first-order PDEs.
Note that, by using the potential $\Psi$, both Eq.~(\ref{DGL3}) as well as
the divergence constraint (\ref{eq:divB}) are
equivalent to Eq.~(\ref{DGL2}), and therefore a solution of Eq.~(\ref{DGL2})
automatically satisfies Eqs.~(\ref{eq:divB}) and (\ref{DGL3}).
Expressing Eqs.~(\ref{DGL5})
and (\ref{DGL6}) in terms of the spherical coordinates introduced with
(\ref{CT1}) as
\begin{eqnarray}
  \partial_r \Psi &=& u_0 \cos{(\vartheta)} B_{\varphi} \\ 
  \partial_{\vartheta} \Psi &=& u_0 \, \left( \frac{q}{r}
  - r \sin{(\vartheta)} \right) B_{\varphi} \ ,
\end{eqnarray}
one finds  
\begin{equation}\label{eqpsi}
  \Psi = u_0 \big[ \mathcal{H}(\varphi) \, a(r,\vartheta)
  + \mathcal{K}(\varphi) \big]
\end{equation}
with $a(r,\vartheta)
:= \sqrt{r^2 \cos^2{(\vartheta)} + 2 q (\sin{(\vartheta)} - 1)}$,
$\mathcal{H}(\varphi)$ determined in Eq.~(\ref{HFCT}), and $\mathcal{K}$ an
undetermined, real-valued function depending solely on the angular variable
$\varphi$. Substituting $\Psi$ into Eq.~(\ref{DGL7}) yields 
\begin{equation} \label{COMPREL}
  \begin{split}
    B_z =& \left( \tan(\vartheta) - \frac{r^2}{q \cos{(\vartheta)}} \right)
    B_{\rho} \\
    &- \frac{r}{q \cos^2{(\vartheta)}} \big[ \partial_{\varphi}
    \mathcal{H}(\varphi) \, a(r,\vartheta) + \partial_{\varphi}
    \mathcal{K}(\varphi) \big] .
  \end{split}
\end{equation}
Together with Eq.~(\ref{DGL2}), this leads to a first-order PDE for
$B_{\rho}$, reading
\begin{eqnarray}
  \mathcal{M}(r, \vartheta, \varphi) &=& 
  \left( \left[ q - r^2 \sin{(\vartheta)} \right] \partial_r
    - r \cos{(\vartheta)} \, \partial_{\vartheta} \right) B_{\rho}
  \nonumber \\
  &&+ \left( \frac{2 q}{r} - 3 r \sin{(\vartheta)} \right) B_{\rho} \ ,
\end{eqnarray}
where the function $\mathcal{M}(r, \vartheta, \varphi)$ is defined by 
\begin{equation}
  \mathcal{M} := 3 \tan{(\vartheta)} \, \big(
    \partial_{\varphi} \mathcal{H}(\varphi) \, a(r,\vartheta)
    + \partial_{\varphi} \mathcal{K}(\varphi) \big) \ .
\end{equation}
Applying the ansatz $B_{\rho}(r, \vartheta, \varphi) =
\mathcal{L}(r, \vartheta, \varphi) \cos(\vartheta)/r^2$, the zeroth-order
derivative term can be eliminated, leaving first-order contributions and an
inhomogeneity
\begin{equation}
  \left( \left[ r \tan (\vartheta)  - \frac{q}{r \cos (\vartheta) } \right]
  \partial_r  + \partial_{\vartheta} \right) \mathcal{L}
  = - \frac{r \mathcal{M}}{\cos^2 (\vartheta) } \ . 
\end{equation}
By means of the transformation $(r, \vartheta) \mapsto (u, v)$ with
\begin{equation}\label{rtheta} 
  \begin{split}
    r &= \frac{\sqrt{2 [\omega_0(v)
        - q \sin (u + \vartheta_0(v)) ]}}{\cos (u + \vartheta_0(v)) } \\
    \vartheta &= u + \vartheta_0(v) \ ,
  \end{split}
\end{equation}
already motivated by Eqs.~(\ref{Bphi}) to (\ref{F2}), (\ref{ODERHO}), and
(\ref{rsol}), one obtains
\begin{equation}
  \begin{split}
    \partial_{u} \mathcal{L} =& - \frac{3 r(u, v)
      \sin{(u + \vartheta_0(v))}}{\cos^3{(u + \vartheta_0(v))}} \\
    &\times \left( \sqrt{2} \, \partial_{\varphi}
      \mathcal{H}(\varphi) \sqrt{\omega_0(v) - q}
      + \partial_{\varphi} \mathcal{K}(\varphi) \right) 
  \end{split}
\end{equation}
which is solved by integration with respect to the variable $u$, giving
\begin{equation} \label{LANS}
  \begin{split}
    \mathcal{L} =& - 3 \, \left( \sqrt{2} \,
      \partial_{\varphi} \mathcal{H}(\varphi) \sqrt{\omega_0(v) - q}
      + \partial_{\varphi} \mathcal{K}(\varphi) \right) \\
    &\times \int{\frac{r(u, v)
        \sin{(u + \vartheta_0(v))}}{\cos^3{(u + \vartheta_0(v))}} \, \dd u}
    + \mathcal{I}(v, \varphi) \ ,
  \end{split}
\end{equation}
where $\mathcal{I}$ is a constant of integration with respect to $u$.
Following the analytical and algebraical manipulations that are provided in
Appendix~\ref{app:cr}, this integral and, in turn, the $\rho$ component of
the ISMF can be expressed in terms of the incomplete elliptic integrals
$F$ and $E$ of first and second kind
\begin{equation} \label{elliptic}
  \begin{split}
    F (x, n) &:=
    \int\limits_0^x \frac{1}{\sqrt{(1 - k^2) (1 - n^2 k^2)}} \, \dd k \\
    E (x, n) &:=
    \int\limits_0^x\sqrt{\frac{1 - n^2 k^2}{1 - k^2}} \, \dd k
  \end{split}
\end{equation}
as
\begin{equation}
  \begin{split} \label{MFRHO}
    B_{\rho} =& \frac{\rho}{r^3} \, \Biggl[\mathcal{I} \left(\frac{\rho^2}{2}
      + \frac{q z}{r}, \varphi \right) \\
    & \quad\quad - \big( \partial_{\varphi} \mathcal{H}(\varphi) \ a(\rho, z)
    + \partial_{\varphi} \mathcal{K}(\varphi) \big) \\
    & \quad\quad \times \left( \frac{q^{3/2}}{a^2} \,
      {\cal T}(\rho, z) + \frac{r^3 + q z}{\rho^2} \right) \Biggr] \ ,
  \end{split}
\end{equation}
where the cylindrical coordinates have been re-substituted and the auxiliary
function 
\begin{equation} \label{aft}
  {\cal T} :=
  \left(2-\frac{1}{\kappa^2} \right) E\laka -
  \left(1-\frac{1}{\kappa^2} \right) F\laka 
\end{equation}
with the quantities
\begin{equation} \label{laka}
  \lambda := \sqrt{1-\frac{a^2}{\rho^2}}
  \  \hbox{,} \quad
  \kappa := \sqrt{1+\frac{a^2}{4 q}}
\end{equation}
has been introduced. Note that the function
\begin{equation} \label{fcta}
  a(\rho,z) = \sqrt{\rho^2 + 2 q \left( \frac{z}{r} - 1 \right)}
\end{equation}
represented in cylindrical coordinates is the same as the one given in
spherical coordinates below Eq.~(\ref{eqpsi}), and furthermore that
$\partial_{\varphi} \mathcal{H}(\varphi)
= - \bigl(\cos{(\varphi)} B_{x 0} + \sin{(\varphi)} B_{y 0}\bigr)$. 
Ensuring the homogeneity of the inward-convecting, undisturbed magnetic
field at infinity, the boundary conditions
\begin{eqnarray}
  \label{eq:lim_bz}
  \lim_{\rho \rightarrow \infty} B_z = \lim_{z \rightarrow \infty} B_z
  &=& B_{z 0} \\
  \label{eq:lim_brho} \nonumber
  \lim_{\rho \rightarrow \infty} B_{\rho} = \lim_{z \rightarrow \infty} B_{\rho} 
  &=& \cos{(\varphi)} B_{x 0} + \sin{(\varphi)} B_{y 0} \\
  &=& - \partial_{\varphi} \mathcal{H}(\varphi)
\end{eqnarray}
are to be imposed.
Using Eq.~(\ref{COMPREL}) together with Eq.~(\ref{MFRHO}) (and observing that
$\lim_{z \rightarrow \infty} {\cal T} = 0$), we get
\begin{eqnarray}
  B_{z 0} &=& \lim_{z \rightarrow \infty} B_z = -\frac{1}{q}
    \lim_{z \rightarrow \infty} {\cal I}
    \left(\frac{\rho^2}{2}+\frac{q \; z}{r}, \varphi \right) \\ \nonumber
    &=& -\frac{1}{q} \ {\cal I} \left( \lim_{z \rightarrow \infty}
      \left[\frac{\rho^2}{2}+\frac{q \; z}{r} \right], \varphi \right)
    = -\frac{1}{q} \ {\cal I} \left(\frac{\rho^2}{2}+q, \varphi \right)
\end{eqnarray}
from the second limit of Eq.~(\ref{eq:lim_bz}), implying that
${\cal I}(p,\varphi) = -q B_{z 0}=$~const.\ for any $\varphi$ and any
real-valued first argument $p$. Moreover, from the second limit of
Eq.~(\ref{eq:lim_brho})
\begin{eqnarray}
  -\partial_{\varphi} {\cal H} &=& \lim_{z \rightarrow \infty} B_{\rho}
  = \lim_{z \rightarrow \infty} \frac{\rho}{z^3} {\cal I}
  -\partial_{\varphi} {\cal H} - \frac{1}{\rho} \, \partial_{\varphi} {\cal K}
\end{eqnarray}
it follows that $\mathcal{K}(\varphi)=$~const. Evidently, both
$\rho \rightarrow \infty$ limits are satisfied as well for these
choices of ${\cal I}$ and ${\cal K}$.
Finally, the magnetic field components read
\begin{widetext}
  \begin{eqnarray}
    B_{\rho}(\rho,\varphi,z) &=& \, - \frac{q \rho}{r^3} \,
    B_{z 0} + \big( \cos{(\varphi)} B_{x 0} + \sin{(\varphi)} B_{y 0} \big)
    \, \left[\frac{q^{3/2} \rho}{r^3 a} \, {\cal T} + \frac{a}{\rho}
      \left(1 + \frac{q z}{r^3}\right)\right] \label{finalr}\\ 
    &&\nonumber\\
    B_{\varphi}(\rho,\varphi,z) &=& \, \frac{\rho}{a}
    \big(- \sin(\varphi) B_{x 0} + \cos(\varphi) B_{y 0} \big)
    \label{finalp} \\ 
    &&\nonumber\\
    B_z(\rho,\varphi,z) &=& \, \left(1 - \frac{q z}{r^3}\right) B_{z 0}
    + \big( \cos{(\varphi)} B_{x 0} + \sin{(\varphi)} B_{y 0} \big)
    \left[\left(\frac{q z}{r^3} - 1\right) \, \frac{\sqrt{q}}{a} \, {\cal T}
      + \frac{q z^2 a}{r^3 \rho^2}\right] \label{finalz}
  \end{eqnarray}
  or, alternatively,
  \begin{eqnarray}
\label{eq:final_bx}
    B_x(x,y,z) &=& \frac{x}{r^3} \left[ (x B_{x 0} + y B_{y 0}) 
      \left( \frac{q^{3/2}}{a \rho} {\cal T}
        + \frac{a}{\rho^3} \left(r^3+q z\right) \right) - q B_{z 0}
    \right] - \frac{y}{a \rho} ( x B_{y 0} - y B_{x 0} )  \\
    B_y(x,y,z) &=& \frac{y}{r^3} \left[ (x B_{x 0} + y B_{y 0}) 
      \left( \frac{q^{3/2}}{a \rho} {\cal T}
        + \frac{a}{\rho^3} \left(r^3+q z\right) \right) - q B_{z 0}
    \right] + \frac{x}{a \rho} ( x B_{y 0} - y B_{x 0} ) \\
    \label{eq:final_bz}
   B_z(x,y,z) &=& \frac{z}{r^3} \left[ (x B_{x 0} + y B_{y 0})
      \left( \frac{q^{3/2}}{a \rho} {\cal T}
        \left[ 1-\frac{r^3}{q z} \right] + \frac{a}{\rho^3} \ q z \right)
      - q B_{z 0} \right] + B_{z 0}
  \end{eqnarray}
  in Cartesian coordinates.
\end{widetext}
These formulas are the central result of the paper. They represent an
analytical solution for the three-dimensionally structured ISMF in the
vicinity of the heliosphere.
As is shown in Appendix~\ref{app:axis}, on the $z$ axis the magnetic field
components (\ref{eq:final_bx}) to (\ref{eq:final_bz}) assume the particularly
simple form
\begin{eqnarray}
  \frac{B_x|_{\rho=0}}{B_{x 0}} &=& \left(1-\frac{q}{z^2}\right)^{-1/2} =
  \frac{B_y|_{\rho=0}}{B_{y 0}} \\
  \frac{B_z|_{\rho=0}}{B_{z 0}} &=& \left(1-\frac{q}{z^2}\right) \ ,
\end{eqnarray}
implying
\begin{equation}
  \label{eq:absB_axis}
  \|{\bf B}\|_{\rho=0} = \sqrt{\frac{B_{x 0}^2 + B_{y 0}^2}{1 - q/z^2}
    + B_{z 0}^2 \left( 1-\frac{q}{z^2}\right)^2} \ .
\end{equation}
Therefore, at the stagnation point $z=\sqrt{q}$ this magnetic field magnitude
with the asymptotic behavior $\sqrt{(B_{x 0}^2 + B_{y 0}^2)/(1 - q/z^2)}$
tends to infinity as expected, while $B_z$ tends to zero.

Before turning to a quantitative analysis and
comparison with self-consistent configurations obtained from numerical MHD
simulations, we present an alternative derivation, which exploits the
concept of a magnetic field frozen into a plasma flow for a more direct,
physically insightful construction of the ISMF.

\subsection{Derivation via the Concept of Frozen-in Fields}
\label{sec:sol_jk}

\begin{figure}
  \begin{center}
    \includegraphics[width=0.4\textwidth]{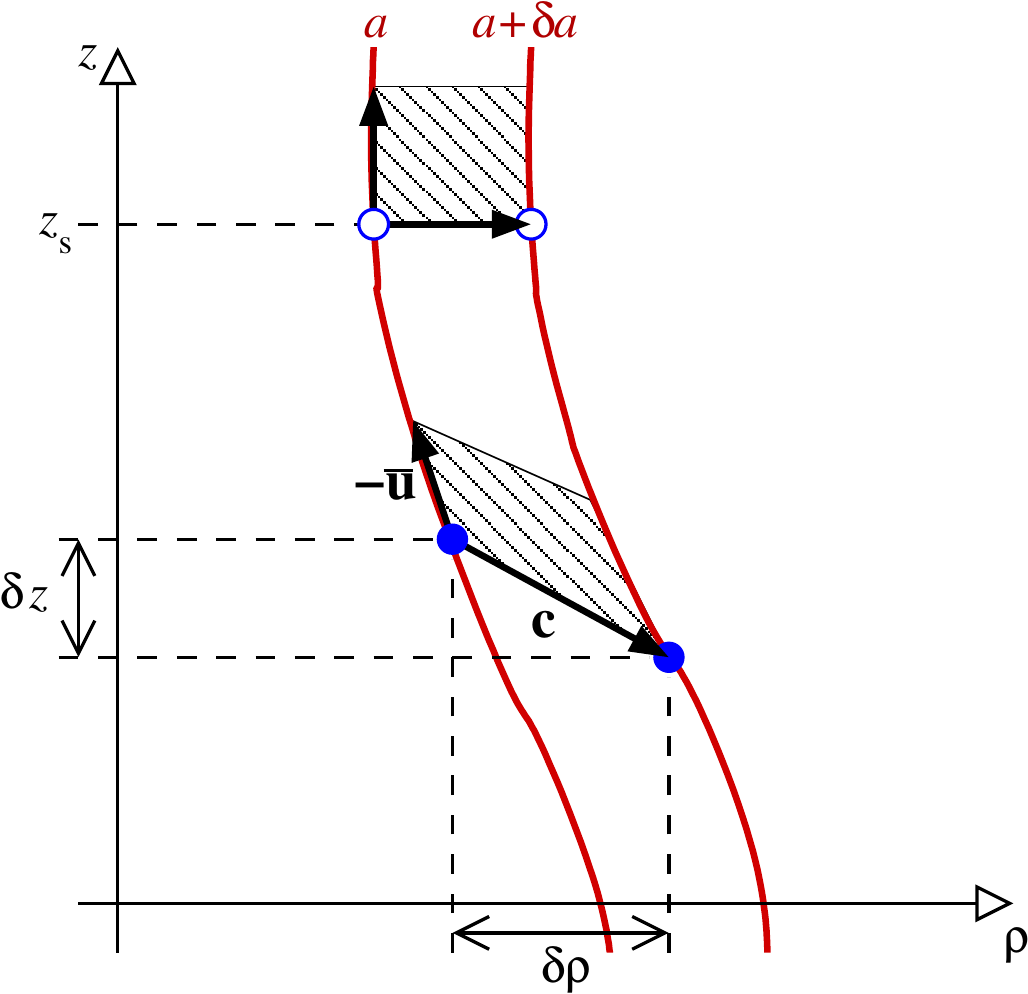}
    \caption{ \label{fig:sketch_voladv}
      Sketch showing the path of P$_1$ [P$_2$] from $z=z_{\rm s}$ (open
      circles) to the new location $(\rho,z)$ [$(\rho+\delta \rho, z+\delta z)$]
      (filled circles) along streamline $a$ [$a+\delta a$].
      The crosshatched areas indicate a co-moving flow parcel whose volume is
      unchanged during transport. Because the flow has a vanishing $\varphi$
      component, this volume is bounded by planes of constant $\varphi$, i.e.,
      its extension perpendicular to the $(\rho,z)$ plane of the paper is
      proportional to $\rho$ (hence the factor $\rho/a$ in the azimuthal base
      vector of ${\cal W}$).
    }
  \end{center}
\end{figure}

\begin{figure}
  \begin{center}
    \includegraphics[width=0.47\textwidth]{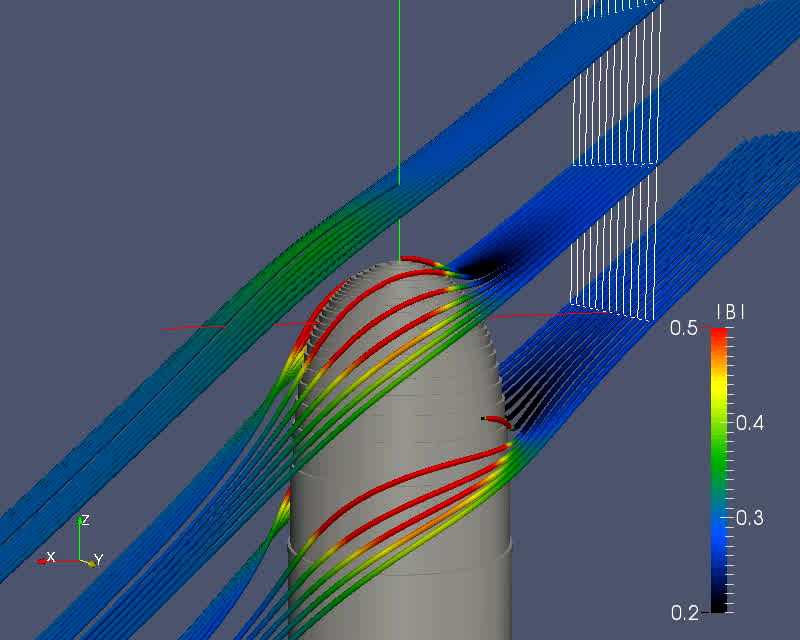}
  \end{center}
  \caption{ \label{fig:render_drape3d}
    Rendering of selected magnetic field lines according to the analytical
    solution (\ref{eq:final_bx}) to (\ref{eq:final_bz}) as they drape around
    the heliopause (grey, solid surface), which is identified
    via Eq.~(\ref{eq:za}) with $a=0$. An MPEG animation visualizing this
    dynamical draping effect is available from the supplementary material.
  }
\end{figure}

The basic idea is to compute the ISMF from the deformation of advected
plasma cells that travel along stream lines, starting in an undistorted
state from a reference location at infinity. In order to
do so, consider two particles P$_{1,2}$ that start at time $t=0$ on adjacent
streamlines $a$ and $\ada$ at the same `height' $z=\zs$. Within a finite time
interval $\Delta t$, P$_1$ travels from $(\rho_a, \zs)$ to $(\rho, z)$, while
P$_2$ travels from $(\rho_{\ada}, \zs)$ to $(\rho+\delta \rho, z+\delta z)$,
where
\begin{equation}
  z_a(\rho_a) = \zs = z_{\ada}(\rho_{\ada})
\end{equation}
with $z_a$ given by Eq.~(\ref{eq:za}). This situation is illustrated in
Fig.~\ref{fig:sketch_voladv}. At $t = \Delta t$, the particles have thus
changed their respective $\rho$ coordinates to $\rho$ and $\rho+\delta\rho$,
such that
\begin{equation}\label{eq:same-tt}
  \int\limits_{\rho_a}^{\rho} \frac{{\rm d} \rp}{\bar{u}_{\rho}(a,\rp)}
  = u_0 \, \Delta t =
  \int\limits_{\rho_{\ada}}^{\rho+\delta \rho}
  \frac{{\rm d} \rp}{\bar{u}_{\rho}(\ada,\rp)}
\end{equation}
holds, where
\begin{equation}
  \begin{split}
    \bar{u}_{\rho}(a,\rho) :=& \frac{q \, \rho}{(\rho^2+z_a(\rho)^2)^{3/2}} \\
    =& \frac{\left[(\rho^2-a^2)(4 \, q +a^2-\rho^2)\right]^{3/2}}{8 \,
      q^2 \rho^2}
  \end{split}
\end{equation}
denotes the $\rho$ component of P$_1$'s flow velocity on the streamline
labeled with $a$, normalized to $u_0$. By definition, the vector
\begin{equation}
 {\bf c} = c_{\rho} {\bf e}_{\rho} + c_z {\bf e}_z :=
 \left(\frac{\delta \rho}{\delta a}\right) {\bf e}_{\rho}
 +            \left(\frac{\delta z}{\delta a}\right) {\bf e}_z \ ,
\end{equation}
pointing from P$_1$ into the direction of P$_2$, is obviously tangential to
the isochrone passing through both points.
Therefore, the set
${\cal W} := \left\{
  {\bf c}, (\rho/a) \ {\bf e}_{\varphi}, -{\bf \bar{u}} \right\}$
defines base vectors that span a non-orthogonal, co-moving coordinate system,
such that the coefficients $(b_1, b_2, b_3)$ of {\bf B} with respect to this
basis remain constant during transport (frozen-in condition). Note that, as
$z \rightarrow \infty$, ${\cal W} \rightarrow
\left\{ {\bf e}_{\rho} , \ {\bf e}_{\varphi},\ {\bf e}_z \right\}$.
Thus, by matching the components of ${\bf B}$ with respect to $\cal{W}$ at
$(\rho, \varphi, z)$ and $(a, \varphi, \infty)$ according to
\begin{eqnarray*}
  {\bf B}|_{z < \infty} &=&
  b_1 {\bf c} + b_2 (\rho/a) \ {\bf e}_{\varphi} + b_3 (-{\bf \bar{u}}) \\
  {\bf B}|_{z \rightarrow \infty} &=&
  b_1 {\bf e}_{\rho} + b_2 {\bf e}_{\varphi} + b_3 {\bf e}_z \\
  &\stackrel{!}{=}& B_{\rho 0} {\bf e}_{\rho}
  + B_{\varphi 0} {\bf e}_{\varphi} + B_{z 0} {\bf e}_z \ ,
\end{eqnarray*}
we obtain $(b_1, b_2, b_3) = ( B_{\rho 0}, B_{\varphi 0}, B_{z 0} )$, and hence
\begin{equation}
  \label{eq:bsol_comp}
  \begin{split}
    {\bf B}(\rho,\varphi,z) =& B_{\rho 0} \ {\bf c}
    + B_{\varphi 0} (\rho/a) \ {\bf e}_{\varphi} - B_{z 0} \ {\bf \bar{u}} \\
    =& \left[ B_{\rho 0} \ c_{\rho} - B_{z 0} \ \bar{u}_{\rho} \right]
    {\bf e}_{\rho} + B_{\varphi 0} (\rho/a) \ {\bf e}_{\varphi} \\
    &+ \left[ B_{\rho 0} \ c_z - B_{z 0} \ \bar{u}_z \right] {\bf e}_z \ .
  \end{split}
\end{equation}
As is shown in Appendix~\ref{app:jk}, the condition of equal travel times
(\ref{eq:same-tt}) can be used to derive explicit expressions for the vector
components $c_{\rho} = \delta \rho/\delta a$ and $c_z = \delta z/\delta a$,
which are the only remaining unknowns in Eq.~(\ref{eq:bsol_comp}).
Using these expressions and the auxiliary variables ${\cal T}$ and $a$
as defined in Eqs.~(\ref{aft}) and (\ref{fcta}), the resulting ISMF becomes
\begin{eqnarray}
  B_{\rho} &=& - B_{z 0} \ \frac{q \, \rho}{r^3} + B_{\rho 0}
  \left[ \frac{q^{3/2} \, \rho}{a \, r^3} {\cal T}
    + \frac{a}{\rho} \left(1+\frac{q \, z}{r^3} \right) \right] \\
  B_{\varphi} &=& B_{\varphi 0} \ \frac{\rho}{a} \\
  B_z &=& B_{z 0} \left(1-\frac{q \, z}{r^3} \right) \nonumber \\
  &&+ B_{\rho 0} 
  \left[\left(\frac{q \, z}{r^3}-1 \right)\! \frac{\sqrt{q}}{a} {\cal T}
    \!+\! \frac{q \, a \, z^2}{\rho^2 r^3} \right]\! 
\end{eqnarray}
which, with the boundary condition of a homogeneous field at infinity
(see Section~\ref{sec:bconds}), i.e.,
\begin{equation}
  \left( \begin{array}{c}
      B_{\rho 0} \\ B_{\varphi 0} \\ B_{z 0}
    \end{array} \right) =
  \left( \begin{array}{c}
      B_{x 0}  \cos (\varphi) + B_{y 0} \sin (\varphi) \\
      -B_{x 0} \sin (\varphi) + B_{y 0} \cos (\varphi) \\ B_{z 0} 
    \end{array} \right) \ ,
\end{equation}
is identical to the representation given in Eqs.~(\ref{finalr}) to
(\ref{finalz}).
The three-dimensional field line geometry is illustrated in
Fig.~\ref{fig:render_drape3d}.

\section{Comparison with Numerical Results}

In order to estimate the degree of accuracy of the field
solution (\ref{finalr}) to (\ref{finalz}), we performed 3D single-fluid MHD
simulations of the LISM--solar wind (SW)-interaction using the {\sc Cronos}
code. For details of the code see \citet{Kissmann-etal-2008} and
\citet{Wiengarten-etal-2014}.
The computational volume covers the region with $(\rho,\varphi,z)
\in [0, 1500]$~AU~$\times \ [0,2\pi] \times [-1500,1000]$~AU
and a grid size of
$N_{\rho} \times N_{\varphi} \times N_z = 150 \times 180 \times 250$,
implying a lateral cell extension of $\Delta \rho = \Delta z = 10$~AU and
an angular cell size of $\Delta \varphi = 2^{\circ}$. The relatively large
extent in the $\rho$ direction was chosen to ensure that the solution is
not contaminated by spurious effects possibly originating at that boundary.
The LISM plasma is incident from the positive $z$ direction, and the ISMF
of strength $\|{\bf B}_0\| = 0.3$~nT is oriented in the
$x$-$z$ plane (i.e., $B_{y 0} = 0$), with an inclination of
$\angle({\bf u},{\bf B})=50^{\circ}$, close to the value of $49^{\circ}$
suggested by \citet{Heerikhuisen-etal-2014}. All other parameters are
identical to those used for the plasma-only case (i.e., not considering
interstellar neutral hydrogen) in the heliospheric benchmark
comparison by \citet{Mueller-etal-2008}, cf.\ Table~1 in that paper.

The full set of time-dependent MHD equations are solved for
$\sim$800~years of physical time until a sufficiently stationary state is
reached. In order to unambiguously determine the heliopause position in the
simulation, an additional equation
\begin{equation}
  \partial_t \psi = - ({\bf u} \cdot \nabla) \psi
\end{equation}
for a passive tracer $\psi({\bf r},t)$ is integrated, with initial condition
\begin{equation}
  \psi({\bf r},0) = \left\{
    \begin{array}{rcl} +1 &:& {\rm SW} \\ -1 &:& {\rm LISM} \ , \end{array}
  \right.
\end{equation}
such that the heliopause can at later times conveniently be identified as
the iso-surface defined by $\psi=0$.
Visual inspection reveals that a satisfactory agreement of the
up- and crosswind heliopause distances along the Cartesian axes of the
simulation volume (except for the $-x$ direction, see comment 1 below) with
the respective
predictions via Eq.~(\ref{eq:za}), i.e., $d_{\rm upwind} = \sqrt{q}$ and
$d_{\rm crossw} = \sqrt{2q}$, is obtained for the choice
$q = (125 \ {\rm AU})^2$, which will thus be used throughout the following
analysis.

Fig.~\ref{fig:compare_cuts} shows a quantitative comparison of the three
magnetic field components along the Cartesian $x$, $y$, and $z$ axes.
In view of the simplifications that have led to the incompressible
steady-state induction equation (\ref{eq:FL}) ---
but not the derivation of the magnetic field resulting from it, which is
exact and void of any additional assumptions or approximations ---, the
agreement is surprisingly satisfactory, save for the following three points:
\begin{enumerate}
\item Along the negative $x$ axis, the agreement is evidently least favorable.
  In this region, which could be called the 'magnetic wake,' the reduced
  magnetic pressure causes a significant outward excursion of the heliopause
  surface. Given that the advecting flow field is axially symmetric and thus
  cannot differentiate between both sides, this excursion is left unaccounted
  for. This region is admittedly a weak spot of our model, which it however
  shares with every other analytical heliosphere shape model that we know of.
\item Close to the heliopause, the field strength of the model necessarily
  tends to infinity, which is of course unphysical. In reality, the field
  strength would grow via pile-up until it becomes dynamically relevant and
  induces a non-linear modification to the flow, which will self-consistently
  settle into a new stationary equilibrium. Furthermore, processes like
  reconnection will prevent infinite magnetic field values, causing the
  field to attain a finite strength just outside the heliopause instead.
  
  It should however be noted that the actual disagreement just outside the
  heliopause is not as large as the right column of
  Fig.~\ref{fig:compare_cuts} would suggest:
  Due to diffusive effects ('numerical resistivity') induced by the finite
  cell size of our simulation, the field strength will tend to the
  corresponding value inside the heliopause, which was chosen to be zero
  here (and in reality would not be zero but in any case much smaller than
  the outside value as well), which is equally unphysical.
  If the resolution was increased considerably (and beyond what our
  resources would permit), spurious diffusive effects can be expected to
  diminish, leading to a more favorable comparison in the spirit of
  Fig.~\ref{fig:compare_cuts}. Therefore, the differences close to the
  heliopause clearly overestimate the actual magnitude of disagreement in
  this respect.
\item As can be seen from the lower two plots of Fig.~\ref{fig:compare_cuts},
  the analytical field solution in the upwind direction remains almost
  indistinguishable from its interstellar value for most of the displayed
  area, showing a moderate ten percent increase in absolute value only at
  a heliocentric distance of $\sim$201~AU according to
  Eq.~(\ref{eq:absB_axis}), and an increase by a factor of two at a mere
  135~AU, i.e., just 10~AU outside the heliopause. Our simulation, on the
  other hand,
  shows the influence of the solar wind's presence to extend over several
  hundred AU in the upwind direction. This discrepancy could be viewed as
  another weakness of our model formulas, although it should be noted that
  the LISM field strength was deliberately chosen high enough to prevent the
  formation of a bow shock.
  Should such a shock exist, it would form at a distance of about 356~AU
  (or 245~AU considering the influence of a neutral particle population)
  according to the heliospheric benchmark by \citet{Mueller-etal-2008}.
  The additional pressure of the ISMF, which was not included in that
  benchmark, would push the bow shock still further inwards.
  From the shock on outwards, all field components would then be
  identical to their respective LISM values beyond the shock. This would
  then again bring them into excellent agreement with our model's prediction.
\end{enumerate}

\section{Summary and Conclusions}

We have derived an analytical formula for the interstellar magnetic field
in the vicinity of the heliosphere under the assumption that a homogeneous
ISM field is being passively advected by an incompressible Rankine-type flow
field, consisting of the superposition of the radial solar wind (as a point
source) and the homogeneous LISM flow. The inclination of the LISM field at
infinity may be chosen freely.
Unlike several previous models for the large-scale heliospheric magnetic
field structure, the one presented here is consistent with a known velocity
field in the sense that both fields together satisfy the stationary induction
equation at any given point.

To derive the explicit formulas for all magnetic vector components, two
complementary approaches were employed, namely a rigorous mathematical
procedure to obtain the solution of the corresponding system of coupled
partial differential equations, and a second approach based on the physical
notion of magnetic elements being kinematically frozen into the prescribed
flow.
The solution thus obtained is exact, i.e., it does not require any additional
assumptions or approximations, and is valid over the entire parameter range
of field strengths and inclination angles.

In order to judge the usefulness of our results for various applications in the
field of heliospheric physics (such as cosmic ray propagation and related
diffusion processes), we performed a quantitative comparison with fully
self-consistent direct numerical MHD simulations, and found very reasonable
agreement, except for the 'magnetic wake' side and the immediate vicinity
of the heliosphere, where our model's field strength necessarily tends to
infinity. However, the affected layer of unphysically high field strength
is rather thin.
Additionally, depending on the nature of a given application, it should
be possible to remove the aforementioned infinities by normalization to a
finite maximum value. The agreement in the upwind direction is more
pronounced in cases where a bow shock is present.
As a further potential application for our field model, it could also be used
as initial condition for investigations employing numerical MHD codes.

In conclusion, the exact analytical solution will be beneficial for studies
of the interaction region of the heliosphere with the LISM comprising
the transport of cosmic rays
\citep[e.g.][]{Scherer-etal-2011, Herbst-etal-2012, Strauss-etal-2013}
and of pick-up ions and energetic neutral atoms
\citep[e.g.][]{Schwadron-McComas-2013, McComas-etal-2014}
in the outer heliosheath, the potential relation of so-called TeV
anisotropies of Galactic cosmic rays to the heliotail
\citep[e.g.][]{Schwadron-etal-2014, Zhang-etal-2014}, and the
characteristics of the magnetized thermal plasma
\citep[e.g.][]{Gurnett-etal-2013, Burlaga-Ness-2014}.

\acknowledgments
\section*{Acknowledgments}

We are grateful to Frederic Effenberger, Ian Lerche, and Klaus Scherer for 
various helpful discussions. We acknowledge financial support via the project 
FI~706/15-1 funded by the Deutsche Forschungsgemeinschaft (DFG). We also 
appreciate discussions at the team meeting `Heliosheath Processes and
Structure of the Heliopause: Modeling Energetic Particles, Cosmic Rays, and
Magnetic Fields' supported by the International Space Science Institute
(ISSI) in Bern, Switzerland.

\begin{widetext}
  
\begin{figure}
  \begin{center}
    \includegraphics[width=\textwidth]{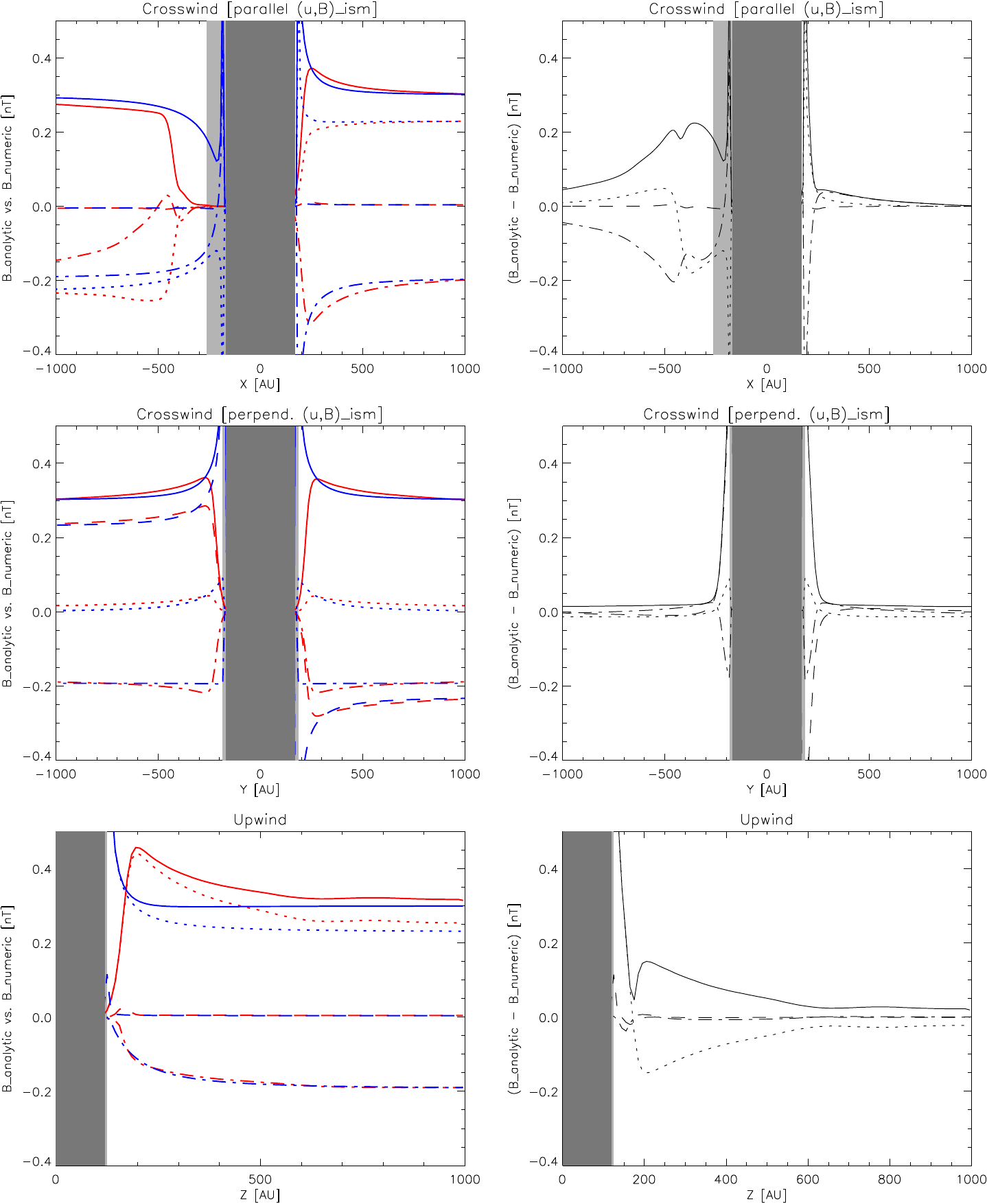}
    \caption{ \label{fig:compare_cuts}
      Left column: Comparison charts of one-dimensional cuts showing
      $B_{\rho}$ (dotted), $B_{\varphi}$ (dashed), $B_z$ (dash-dotted), and
      $\|{\bf B}\|$ (solid) of the analytical solution
      (\ref{finalr}) to (\ref{finalz}) (blue) versus the numerical MHD
      results (red) along all three Cartesian axes.
      Right column: Same plots showing only the respective differences
      (blue minus red).
      The area shaded in light gray marks the heliopause interior according
      to the numerical value of the tracer $\psi$, whereas the area blocked
      out in dark gray indicates the heliopause interior as defined by
      $\{(\rho,z) | z \le z_0(\rho) \}$ (cf.\ Eq.~(\ref{eq:za})), which is
      not subject of the present study, and for which our analytical solution
      is not valid.
      The numerical $\|{\bf B}\|$ solution at the upwind
      boundary (located at $z=1000$~AU) is slightly above 0.3~nT due to
      the fact that the entire upwind region is still sub-Alfv\'enic, and
      thus in the absence of a bow shock allows the heliosphere's influence
      to propagate all the way to that boundary.
    }
  \end{center}
\end{figure}

\appendix

\section{A. Evaluation of the Integral in
  Eq.~(\ref{LANS})}
\label{app:cr}
\numberwithin{equation}{section}

The integral in Eq.~(\ref{LANS}) can be re-written as follows.
First, since the variable $v$ is treated as a constant in the integration
with respect to $u$, one can formulate the integral in terms of the new
integration variable
$\zeta = u + \vartheta_0(v) \in \left[- \pi/2, \pi/2\right]$, yielding
\begin{equation}\label{inint}
  \int{\frac{r(u, v)
      \sin{(u + \vartheta_0(v))}}{\cos^3{(u + \vartheta_0(v))}} \,
    \dd u} = \sqrt{2 \omega_0(v)} \int \frac{\sin{(\zeta)}}{\cos^4{(\zeta)}}
  \sqrt{1 - \tau \sin{(\zeta)}} \, \dd \zeta \ ,
\end{equation}
where $\tau := q/\omega_0(v)$ and $r(u, v)$ is given in Eq.~(\ref{rtheta}).
Twofold integration by parts leads to 
\begin{equation}\label{TFIBP}
  \int \frac{\sin{(\zeta)}}{\cos^4{(\zeta)}} \sqrt{1 - \tau \sin{(\zeta)}} \,
  \dd \zeta = \frac{\sqrt{1 - \tau \sin{(\zeta)}}}{3 \cos^3{(\zeta)}}
  + \frac{\tau \tan{(\zeta)}}{6 \sqrt{1 - \tau \sin{(\zeta)}}}
  - \frac{\tau^2}{12}
  \int \frac{\sin{(\zeta)}}{(1 - \tau \sin{(\zeta)})^{3/2}} \dd \zeta
\end{equation}
in which the integral on the right hand side can be re-written as
\begin{equation}\label{Irhs}
  \begin{split}
    \int \frac{\sin (\zeta) }{(1 - \tau \sin (\zeta) )^{3/2}} \,\dd\zeta &
    = \int
    \frac{1 - (1 - \tau \sin (\zeta) )}{\tau (1 - \tau \sin{(\zeta)})^{3/2}}
    \,\dd\zeta \\ 
    &= \frac{1}{\tau} \int \frac{1}{(1 - \tau \sin{(\zeta)})^{3/2}} \,\dd\zeta 
    - \frac{1}{\tau} \int \frac{1}{\sqrt{1 - \tau \sin{(\zeta)}}}
    \,\dd\zeta \ .
  \end{split}
\end{equation}
With the identity $1 - \tau \sin{(\zeta)} =
(1 - \tau) \bigl(1 + w \sin^2{(\zeta/2 - \pi/4)}\bigr)$, where
$w := 2 \tau/(1 - \tau)$, and the substitution
$m = \sin{(\zeta/2 - \pi/4)} \in [-1, 0]$, Eq.~(\ref{Irhs}) becomes
\begin{equation}\label{EEEF}
  \begin{split}
    \frac{1}{\tau} \int \frac{1}{(1 - \tau \sin (\zeta) )^{3/2}} \,\dd\zeta
    - \frac{1}{\tau} \int \frac{1}{\sqrt{1 - \tau \sin (\zeta) }}
    \,\dd\zeta =& \frac{2}{\tau (1-\tau)^{3/2}}\int\frac{1}{\sqrt{1-m^2}
      \, (1 + w \, m^2)^{3/2}} \,\dd m \\ 
    &- \frac{2}{\tau \sqrt{1 - \tau}} \int \frac{1}{\sqrt{1 - m^2} \,
      \sqrt{1 + w \, m^2}} \,\dd m \ .
  \end{split}
\end{equation}
The first integral on the right hand side can be brought into the following
form
\begin{equation}
  \begin{split}
    \int \frac{1}{\sqrt{1 - m^2} \, (1 + w \, m^2)^{3/2}} \,\dd m & =
    \int \frac{\sqrt{1 + w \, m^2}}{\sqrt{1 - m^2} \, (1 + w m^2)^{2}} \,\dd m
    = \int \sqrt{1 + w \, m^2} \,\, \frac{1 - 2 m^2 - w \, m^2 + 2 m^2 
      + w \, m^2}{\sqrt{1 - m^2} \, (1 + w \, m^2)^{2}} \,\dd m \\
    & = \int \sqrt{1 + w \, m^2} \ \frac{\dd}{\dd m}
    \biggl(\frac{m \sqrt{1 - m^2}}{1 + w \, m^2}\biggr) \,\dd m 
    + \int \frac{(2 + w) m^2}{\sqrt{1 - m^2} \, (1 + w \, m^2)^{3/2}}
    \,\dd m \ .
  \end{split}
\end{equation}
Then, integrating by parts, one obtains  
\begin{equation}
  \int \frac{1}{\sqrt{1 - m^2} \, (1 + w \, m^2)^{3/2}} \,\dd m
  = \frac{m \sqrt{1 - m^2}}{\sqrt{1 + w \, m^2}} + \frac{1}{w} 
  \int \sqrt{\frac{1 + w \, m^2}{1 - m^2}} \,\dd m
  - \frac{1}{w} \int \frac{1}{\sqrt{1 - m^2} \, (1 + w \, m^2)^{3/2}} \,\dd m
\end{equation}
and hence
\begin{equation}\label{EE}
\int \frac{1}{\sqrt{1 - m^2} \, (1 + w \, m^2)^{3/2}} \,\dd m
 = \frac{w \, m}{1 + w} \, \sqrt{\frac{1 - m^2}{1 + w \, m^2}} 
 + \frac{1}{1 + w} \int \sqrt{\frac{1 + w \, m^2}{1 - m^2}} \,\dd m \ .
\end{equation}
By means of the incomplete elliptic integrals of the first and second kind,
$F$ and $E$ defined in Eq.~(\ref{elliptic}), the initial integral
(\ref{inint}) can be
given, subsequently substituting (\ref{EE}) into (\ref{EEEF}), (\ref{EEEF})
into (\ref{Irhs}), and (\ref{Irhs}) into (\ref{TFIBP}), in the following form
\begin{equation} \label{eq:int_fe}
  \begin{split}
    \int{\frac{r(u, v)
        \sin (u + \vartheta_0(v)) }{\cos^3 (u + \vartheta_0(v)) } \,\dd u}
    =& \frac{r(u, v)}{3 \cos^2{(u + \vartheta_0(v))}} \,
    \biggl(1 + \frac{q \sin{(u + \vartheta_0(v))}}{r^2(u, v)}\biggr)
    + \frac{q^2 \omega_0(v)}{3 \, r(u, v) [\omega_0(v)^2 - q^2]} \\ 
    +& \frac{q}{3 \sqrt{2 (\omega_0(v) - q)}} \left[ F\bigl(s, \, t\bigr)
    - \frac{\omega_0(v)}{\omega_0(v) + q} \, E\bigl(s, \, t\bigr) \right] \ ,
  \end{split}
\end{equation}
where 
\begin{equation}
  \begin{split}
    &s := \sin{\left(\frac{u + \vartheta_0(v)}{2} - \frac{\pi}{4}\right)} \\
    &t := \frac{2 \sqrt{q} \ \ii}{\sqrt{r^2(u, v) \cos^2{(u + \vartheta_0(v))}
        + 2 q [\sin{(u + \vartheta_0(v))} - 1]}} \in \mathbb{C} \ .
  \end{split}
\end{equation}
Using the transformation formulas
\begin{eqnarray*}
  F \left( x, \, \ii n \right) &=& \frac{1}{n} \,
  F \left( \frac{n \, x}{\sqrt{1 + n^2 \, x^2}}, \,
    \frac{\sqrt{1 + n^2}}{n} \right) \\ \\
  E \left( x, \, \ii n \right) &=& \frac{1}{n} \,
  F \left( \frac{n \, x}{\sqrt{1 + n^2 \, x^2}}, \,
    \frac{\sqrt{1 + n^2}}{n} \right)
  + n \, E \left( \frac{n \, x}{\sqrt{1 + n^2 \, x^2}}, \,
    \frac{\sqrt{1 + n^2}}{n} \right)
  - n^2 x \, \sqrt{\frac{1 - x^2}{1 + n^2 \, x^2}} \ ,
\end{eqnarray*}
the elliptic integrals $F(s, t)$ and $E(s, t)$ in Eq.~(\ref{eq:int_fe})
can be expressed in terms of the real-valued arguments $\lambda$ and
$\kappa$ as defined in Eq.~(\ref{laka}), yielding
\begin{eqnarray}
  F (s,\, t) &=& F \left(- \frac{1}{2} \, \sqrt{\frac{\rho^2 - a^2}{q}},\,
    \frac{2 \sqrt{q}}{a} \, \ii \right) \
  = - \frac{a}{2\sqrt{q}} \, F\left(\lambda,\, \kappa\right)
\\
  E (s,\, t) &=& E\left(- \frac{1}{2} \, \sqrt{\frac{\rho^2 - a^2}{q}}, \,
    \frac{2 \sqrt{q}}{a} \, \ii \right) = \frac{\sqrt{\rho^2 - a^2}
    \sqrt{4q + a^2 - \rho^2}}{\rho \, a} - \frac{2 \sqrt{q}}{a}
  \left[ \frac{a^2}{4 q} F\left(\lambda,\, \kappa\right)
    + E\left(\lambda,\, \kappa\right) \right] \ .
\end{eqnarray}
Moreover, one obtains
\begin{equation}
  F \left(s, \, t\right) - \frac{\omega_0(v)}{\omega_0(v) + q} \,
  E \left(s, \, t\right) = \frac{\sqrt{q}}{a} \, {\cal T}
  - \frac{\sqrt{\rho^2 - a^2} \sqrt{4 q + a^2 - \rho^2}}{\rho \, a} \,
  \frac{a^2 + 2 q}{a^2 + 4 q}
\end{equation}
for the square brackets in Eq.~(\ref{eq:int_fe}), with ${\cal T}$ defined in
Eq.~(\ref{aft}). Then, the initial integral (\ref{inint}) becomes
\begin{equation}
  \int{\frac{r(u, v) \sin (u + \vartheta_0(v))}{\cos^3(u + \vartheta_0(v))}
    \,\dd u} \ = \frac{1}{3} \left( \frac{q^{3/2}}{a^2} \, {\cal T}
    + \frac{r^3 + q z}{\rho^2} \right) \ .
\end{equation}
~\\
\section{B. Explicit Derivation of the Components
  of Vector {\bf c}}
\label{app:jk}

\numberwithin{equation}{section}
In order to derive explicit formulas for $\delta \rho/\delta a$ and
$\delta z/\delta a$, we first need to find the integral of $1/\bar{u}_{\rho}$
with respect to $\rho$, see Eq.~(\ref{eq:same-tt}), which can be
expressed using $\lambda$ and $\kappa$ defined in Eq.~(\ref{laka}) as
\begin{equation}
  H := \int \frac{\dd \rho}{\bar{u}_{\rho}(a,\rho)}
  = \int \frac{8 q^2 \rho^2}{[(\rho^2-a^2) (4q+a^2-\rho^2)]^{3/2}} \,\dd \rho
  = \sqrt{q} \int \frac{\sqrt{1-\lambda^2}}{\lambda^2}
  \frac{1}{(1-\kappa^2\lambda^2)^{3/2}} \,\dd \lambda \ .
\end{equation}
Note that, since the integration occurs along a fixed streamline, both $a$
and $\kappa$ are to be treated as constants.
Multiplication of the integrand by
$1=[1-(\kappa\lambda)^2]+(\kappa\lambda)^2$ gives
\begin{equation}
  \begin{split}
    \frac{H}{\sqrt{q}} &= \int \frac{\sqrt{1-\lambda^2}}{\lambda^2}
    \frac{1}{\sqrt{1-\kappa^2\lambda^2}} \,\dd \lambda +
    \kappa^2 \int \frac{\sqrt{1-\lambda^2}}{
      (1-\kappa^2\lambda^2)^{3/2}} \,\dd \lambda \\
    &= -\int \sqrt{1-\lambda^2} \ \frac{\dd}{\dd \lambda}
    \bigg( \frac{\sqrt{1-\kappa^2\lambda^2}}{\lambda} \bigg) \,\dd \lambda
    + \kappa^2 \int \sqrt{1-\lambda^2} \ \frac{\dd}{\dd \lambda}
    \bigg( \frac{\lambda}{\sqrt{1-\kappa^2\lambda^2}} \bigg) \,\dd \lambda \ ,
  \end{split}
\end{equation}
which, via integration by parts, yields
\begin{equation}
  \begin{split}
    \frac{H}{\sqrt{q}} &= -\bigg( \sqrt{1-\lambda^2} \
    \frac{\sqrt{1-\kappa^2\lambda^2}}{\lambda}
    + \underbrace{\int \frac{\sqrt{1-\kappa^2\lambda^2}}{\sqrt{1-\lambda^2}}
      \,\dd \lambda}_{=E\laka} \bigg)
    + \bigg( \kappa^2 \lambda
    \frac{\sqrt{1-\lambda^2}}{\sqrt{1-\kappa^2\lambda^2}}
    + \underbrace{\int
      \frac{\kappa^2 \lambda^2}{\sqrt{1-\lambda^2}\sqrt{1-\kappa^2\lambda^2}}
      \,\dd\lambda  }_{=F\laka-E\laka} \bigg) \\
    &= F\laka - 2 E\laka - \frac{1-2 \kappa^2 \lambda^2}{\lambda}
    \sqrt{\frac{1-\lambda^2}{1-\kappa^2 \lambda^2}} \ .
  \end{split}
\end{equation}
Re-substituting the original arguments $a$ and $\rho$, we write
$H$ as $H(a,\rho) = G(a,\rho) - z_a(\rho)$, where
\begin{equation}
  G(a,\rho) := \sqrt{q} \left[
    F \left( \sqrt{1-\frac{a^2}{\rho^2}},\sqrt{1+\frac{a^2}{4 q}} \right)
    -2 \, E \left( \sqrt{1-\frac{a^2}{\rho^2}},\sqrt{1+\frac{a^2}{4 q}} \right)
  \right] + \frac{2 q}{\sqrt{\rho^2+z_a(\rho)^2}} \ .
\end{equation}
Condition (\ref{eq:same-tt}) for equal travel times thus becomes
\begin{eqnarray*}
  H(a,\rho)- H(a,\rho_a) &=&
  H(\ada,\rho+\delta\rho)- H(\ada,\rho_{\ada}) \\
  \Rightarrow \quad H(\ada,\rho_{\ada}) - H(a,\rho_a) &=&
  \partial_a H(a,\rho) \ \delta a +
  \partial_{\rho} H(a,\rho) \ \delta \rho + {\cal O}(\delta^2) \ .
\end{eqnarray*}
Neglecting terms ${\cal O}(\delta^2)$, one obtains
\begin{equation}
  \big[ G(\ada,\rho_{\ada}) - \underbrace{z_{\ada}(\rho_{\ada})}_{=\zs} \big] -
  \big[ G(   a,\rho_{   a}) - \underbrace{z_a    (\rho_a     )}_{=\zs} \big]
  = \partial_a H(a,\rho) \ \delta a
  + \left[1/\bar{u}_{\rho}(a,\rho)\right] \ \delta \rho \ .
\end{equation}
We now consider the limit $\zs \rightarrow \infty$, in which
$\rho_a \rightarrow a$ and $\rho_{\ada} \rightarrow \ada$. Then the left hand
side vanishes due to
\begin{equation}
  \lim_{z_{\rm s}\rightarrow \infty}
  \big( G(\ada,\rho_{\ada}) - G(a,\rho_a) \big) = G(\ada,\ada) - G(a,a)
  \approx \partial_a \underbrace{G(a,a)}_{=0} \ \delta a = 0 \ ,
\end{equation}
while the right hand side remains unaffected by this limit.
This leads to
\begin{eqnarray}
  \nonumber\frac{\delta \rho}{\delta a} &=&
  - \bar{u}_{\rho}(a,\rho) \ \frac{\partial}{\partial a} H(a,\rho) \\
  \nonumber  &=& - \frac{q \, \rho}{r^3} \left[ \sqrt{q} \,
    \frac{\partial}{\partial a}
    \left[ F\laka - 2 \, E\laka \right]
    - \frac{\partial}{\partial a}
    \left( \frac{2 q}{\sqrt{\rho^2+z_a(\rho)^2}} - z_a(\rho) \right) \right]\\
  &=&   \frac{q^{3/2} \; \rho}{a \, r^3} \bigg[ \underbrace{
    \frac{2a^2+4q}{a^2+4q}   \; E\laka -
    \frac{a^2    }{a^2+4q}   \; F\laka}_{={\cal T}} \bigg]
  + \frac{a}{\rho} \left( 1 + \frac{q \, z}{r^3} \right) \ .
\end{eqnarray}
With
\begin{equation}
  c_0(a,\rho) := \frac{\partial z_a(\rho)}{\partial a} =
  \frac{8q^2 \ a \rho}{\left[ (\rho^2-a^2) (4q+a^2-\rho^2) \right]^{3/2}}
  = \frac{a \, r^3}{q \, \rho^2} \ ,
\end{equation}
we furthermore obtain
\begin{eqnarray}
  \frac{\delta z}{\delta a} &=&
  \underbrace{\frac{\partial z}{\partial a}}_{=c_0} +
  \underbrace{\frac{\partial z}{\partial \rho}}_{=u_z/u_{\rho}}
  \frac{\delta \rho}{\delta a} 
  = \frac{ar^3}{q \; \rho^2} + \frac{q \; z/r^3-1}{q \; \rho/r^3}
  \left[ \frac{q^{3/2} \; \rho}{a \; r^3} {\cal T}
    + \frac{a}{\rho} \left( 1+\frac{q \; z}{r^3} \right) \right]
  = \frac{\sqrt{q}}{a} \left( \frac{q \; z}{r^3}-1 \right) {\cal T}
  + \frac{q \; a \; z^2}{\rho^2 r^3} \ .
\end{eqnarray}
These are the desired expressions for $\delta \rho/\delta a$ and
$\delta z/\delta a$ required for the computation of the components of
${\bf c}$ in Eq.~(\ref{eq:bsol_comp}).
~\\
\section{C. The Magnetic Field on the inflow axis}
\label{app:axis}

The Taylor expansions of the functions $a$, $\lambda$, and $\kappa$ given in
Eqs.~(\ref{fcta}) and (\ref{laka}), respectively, at $\rho=0$ are given by
\begin{eqnarray}
a &=& \rho \ \sqrt{1-\frac{q}{z^2}} + {\cal O}(\rho^2) \\
\lambda &=& \frac{\sqrt{q}}{z} + {\cal O}(\rho^2) \\
\kappa &=& 1 + {\cal O}(\rho^2)
\end{eqnarray}
for all relevant values of $\rho$ and $z$. Using these expressions,
the function ${\cal T}$ yields in the limit $\rho \rightarrow 0$
\begin{equation}
  \begin{split}
    \lim_{\rho\rightarrow 0} {\cal T} &=
    \lim_{\rho\rightarrow 0} \left[
      \left(2-\frac{1}{\kappa^2} \right) \int\limits_0^{\lambda}
      \sqrt{\frac{1-\kappa^2 k^2}{1-k^2}} \, \dd k -
      \left(1-\frac{1}{\kappa^2} \right)
      \int\limits_0^{\lambda} \frac{1}{\sqrt{(1- k^2)(1-\kappa^2 k^2)}} \,\dd k
    \right] \\
    &= \lim_{\rho\rightarrow 0} \left[
      \left(1+{\cal O}(\rho^2)\right) \int\limits_0^{\lambda} \, \dd k -
      {\cal O}(\rho^2) \int\limits_0^{\lambda} \frac{1}{1- k^2} \,\dd k
    \right] = \lim_{\rho\rightarrow 0} \lambda = \frac{\sqrt{q}}{z} \ .
  \end{split}
\end{equation}
Consequently, on the $z$ axis, one obtains for the magnetic fields components
(\ref{finalr}) to (\ref{finalz})
\begin{eqnarray}
  B_{\rho}|_{\rho=0} &=&
  \big( \cos(\varphi) B_{x 0} + \sin(\varphi) B_{y 0} \big)
  \underbrace{\lim_{\rho\rightarrow 0} \left[ \frac{q^2 \rho}{z^4 a} + 
      \frac{a}{\rho} \left(1+\frac{q}{z^2} \right) \right]}_{=(1-q/z^2)^{-1/2}} \\
  B_{\varphi}|_{\rho=0} &=& \left(1-\frac{q}{z^2}\right)^{-1/2}
  \big( -\sin(\varphi) B_{x 0} + \cos(\varphi) B_{y 0} \big) \\
 \label{eq:Bz_axis}   B_z|_{\rho=0} &=& B_{z 0} \left(1-\frac{q}{z^2}\right) +
  \big( \cos(\varphi) B_{x 0} + \sin(\varphi) B_{y 0} \big)
  \underbrace{\lim_{\rho\rightarrow 0} \left[ \left(\frac{q}{z^2}-1 \right)
      \frac{q}{a z} +  \frac{q a}{\rho^2 z} \right]}_{=0} \ ,
\end{eqnarray}
implying the Cartesian components
\begin{eqnarray}
   \label{eq:Bx_axis} B_x|_{\rho=0} &=& \cos(\varphi) B_{\rho}|_{\rho=0} -
  \sin(\varphi) B_{\varphi}|_{\rho=0}
  = B_{x 0} \left( 1-\frac{q}{z^2} \right)^{-1/2} \\
   \label{eq:By_axis} B_y|_{\rho=0} &=& \sin(\varphi) B_{\rho}|_{\rho=0} +
  \cos(\varphi) B_{\varphi}|_{\rho=0}
  = B_{y 0} \left( 1-\frac{q}{z^2} \right)^{-1/2} .
\end{eqnarray}

Alternatively, this result can be obtained more easily by substituting
$\rho=0$ into the original PDEs (\ref{DGL1}) to (\ref{DGL3}), which
then simplify considerably to
\begin{equation}
  z \left(1-\frac{z^2}{q}\right) \partial_z B_{\varphi} = B_{\varphi} \ , \quad
  z \left(1-\frac{z^2}{q}\right) \partial_z B_{\rho}    = B_{\rho} \ , \quad
  z \left(1-\frac{z^2}{q}\right) \partial_z B_z        = -2 B_z \ ,
\end{equation}
and may be solved straightforwardly in this form.

Note that the axis solution (\ref{eq:Bz_axis}) to (\ref{eq:By_axis}) is
consistent with both the notion of {\bf B} being frozen into a co-moving
brick-shaped volume whose side lengths $(L_x, L_y, L_z )$ are proportional
to $(B_x, B_y, B_z)$, implying
\begin{equation}
  \frac{B_z|_{\rho=0}}{B_{z 0}} = \left. \frac{u_z}{u_{z 0}} \right|_{\rho=0} =
  \left. \frac{u_0(q z/r^3-1)}{-u_0} \right|_{\rho=0} = 1- \frac{q}{z^2} \ ,
\end{equation}
as well as with the incompressibility of the advecting flow {\bf u}, from
which it follows that
\begin{equation}
  \frac{B_x|_{\rho=0}}{B_{x 0}} =
  \left(\frac{B_z|_{\rho=0}}{B_{z 0}} \right)^{-1/2}  =
  \frac{B_y|_{\rho=0}}{B_{y 0}}
\end{equation}
because the volume $L_x L_y L_z \sim B_x B_y B_z$ is conserved during the
transport, and the flow is symmetric in $x \leftrightarrow y$.

\end{widetext}
%
%\begin{thebibliography}{}
%\end{thebibliography}
%
\bibliographystyle{apj}
\bibliography{rkf_ismf_revised}

\begin{thebibliography}{}
\expandafter\ifx\csname natexlab\endcsname\relax\def\natexlab#1{#1}\fi

\bibitem[{{Amenomori} \& {Tibet As{$\gamma$}
  Collaboration}(2010)}]{Amenomori-etal-2010}
{Amenomori}, M., \& {Tibet As{$\gamma$} Collaboration}. 2010, Astrophysics and
  Space Sciences Transactions, 6, 49

\bibitem[{{Amenomori} {et~al.}(2006){Amenomori}, {Ayabe}, {Bi}, {Chen}, {Cui},
  {Danzengluobu}, {Ding}, {Ding}, {Feng}, {Feng}, {Feng}, {Gao}, {Geng}, {Guo},
  {He}, {He}, {Hibino}, {Hotta}, {Hu}, {Hu}, {Huang}, {Huang}, {Jia}, {Kajino},
  {Kasahara}, {Katayose}, {Kato}, {Kawata}, {Labaciren}, {Le}, {Li}, {Li},
  {Lou}, {Lu}, {Lu}, {Meng}, {Mizutani}, {Mu}, {Munakata}, {Nagai}, {Nanjo},
  {Nishizawa}, {Ohnishi}, {Ohta}, {Onuma}, {Ouchi}, {Ozawa}, {Ren}, {Saito},
  {Saito}, {Sakata}, {Sako}, {Sasaki}, {Shibata}, {Shiomi}, {Shirai},
  {Sugimoto}, {Takita}, {Tan}, {Tateyama}, {Torii}, {Tsuchiya}, {Udo}, {Wang},
  {Wang}, {Wang}, {Wang}, {Wu}, {Xue}, {Yamamoto}, {Yan}, {Yang}, {Yasue},
  {Ye}, {Yu}, {Yuan}, {Yuda}, {Zhang}, {Zhang}, {Zhang}, {Zhang}, {Zhang},
  {Zhang}, {Zhaxisangzhu}, {Zhou}, \& {Tibet AS{$\gamma$}
  Collaboration}}]{Amenomori-etal-2006}
{Amenomori}, M., {Ayabe}, S., {Bi}, X.~J., {et~al.} 2006, Science, 314, 439

\bibitem[{{Belcher} {et~al.}(1993){Belcher}, {Lazarus}, {McNutt}, \&
  {Gordon}}]{Belcher-etal-1993}
{Belcher}, J.~W., {Lazarus}, A.~J., {McNutt}, Jr., R.~L., \& {Gordon}, Jr.,
  G.~S. 1993, \jgr, 98, 15177

\bibitem[{{Ben-Jaffel} {et~al.}(2013){Ben-Jaffel}, {Strumik}, {Ratkiewicz}, \&
  {Grygorczuk}}]{BenJaffel-etal-2013}
{Ben-Jaffel}, L., {Strumik}, M., {Ratkiewicz}, R., \& {Grygorczuk}, J. 2013,
  \apj, 779, 130

\bibitem[{{Borovikov} \& {Pogorelov}(2014)}]{Borovikov-Pogorelov-2014}
{Borovikov}, S.~N., \& {Pogorelov}, N.~V. 2014, \apjl, 783, L16

\bibitem[{{Burlaga} \& {Ness}(2014)}]{Burlaga-Ness-2014}
{Burlaga}, L.~F., \& {Ness}, N.~F. 2014, \apj, 784, 146

\bibitem[{{Desiati} \& {Lazarian}(2013)}]{Desiati-Lazarian-2013}
{Desiati}, P., \& {Lazarian}, A. 2013, \apj, 762, 44

\bibitem[{{Frisch}(2007)}]{Frisch-2007}
{Frisch}, P.~C. 2007, \ssr, 130, 355

\bibitem[{Gurnett {et~al.}(2013)Gurnett, Kurth, Burlaga, \&
  Ness}]{Gurnett-etal-2013}
Gurnett, D., Kurth, W., Burlaga, L., \& Ness, N. 2013, Science, 341, 1489

\bibitem[{{Heerikhuisen} {et~al.}(2008){Heerikhuisen}, {Pogorelov},
  {Florinski}, {Zank}, \& {le Roux}}]{Heerikhuisen-etal-2008}
{Heerikhuisen}, J., {Pogorelov}, N.~V., {Florinski}, V., {Zank}, G.~P., \& {le
  Roux}, J.~A. 2008, \apj, 682, 679

\bibitem[{{Heerikhuisen} {et~al.}(2014){Heerikhuisen}, {Zirnstein}, {Funsten},
  {Pogorelov}, \& {Zank}}]{Heerikhuisen-etal-2014}
{Heerikhuisen}, J., {Zirnstein}, E.~J., {Funsten}, H.~O., {Pogorelov}, N.~V.,
  \& {Zank}, G.~P. 2014, \apj, 784, 73

\bibitem[{{Herbst} {et~al.}(2012){Herbst}, {Heber}, {Kopp}, {Sternal}, \&
  {Steinhilber}}]{Herbst-etal-2012}
{Herbst}, K., {Heber}, B., {Kopp}, A., {Sternal}, O., \& {Steinhilber}, F.
  2012, \apj, 761, 17

\bibitem[{{Izmodenov} {et~al.}(2005){Izmodenov}, {Alexashov}, \&
  {Myasnikov}}]{Izmodenov-etal-2005b}
{Izmodenov}, V., {Alexashov}, D., \& {Myasnikov}, A. 2005, \aap, 437, L35

\bibitem[{{Kissmann} {et~al.}(2008){Kissmann}, {Kleimann}, {Fichtner}, \&
  {Grauer}}]{Kissmann-etal-2008}
{Kissmann}, R., {Kleimann}, J., {Fichtner}, H., \& {Grauer}, R. 2008, \mnras,
  391, 1577

\bibitem[{{McComas} {et~al.}(2014){McComas}, {Lewis}, \&
  {Schwadron}}]{McComas-etal-2014}
{McComas}, D., {Lewis}, W., \& {Schwadron}, N. 2014, Rev.\ Geophys.\, 52,
  doi:10.1002/2013RG000438

\bibitem[{{McComas} {et~al.}(2012{\natexlab{a}}){McComas}, {Dayeh},
  {Allegrini}, {Bzowski}, {DeMajistre}, {Fujiki}, {Funsten}, {Fuselier},
  {Gruntman}, {Janzen}, {Kubiak}, {Kucharek}, {Livadiotis}, {M{\"o}bius},
  {Reisenfeld}, {Reno}, {Schwadron}, {Sok{\'o}{\l}}, \&
  {Tokumaru}}]{McComas-etal-2012}
{McComas}, D.~J., {Dayeh}, M.~A., {Allegrini}, F., {et~al.} 2012{\natexlab{a}},
  \apjs, 203, 1

\bibitem[{{McComas} {et~al.}(2012{\natexlab{b}}){McComas}, {Alexashov},
  {Bzowski}, {Fahr}, {Heerikhuisen}, {Izmodenov}, {Lee}, {M{\"o}bius},
  {Pogorelov}, {Schwadron}, \& {Zank}}]{McComas-etal-2012b}
{McComas}, D.~J., {Alexashov}, D., {Bzowski}, M., {et~al.} 2012{\natexlab{b}},
  Science, 336, 1291

\bibitem[{{Mitchell} {et~al.}(2008){Mitchell}, {Cairns}, {Pogorelov}, \&
  {Zank}}]{Mitchell-etal-2008}
{Mitchell}, J.~J., {Cairns}, I.~H., {Pogorelov}, N.~V., \& {Zank}, G.~P. 2008,
  Journal of Geophysical Research (Space Physics), 113, 4102

\bibitem[{{M{\"u}ller} {et~al.}(2008){M{\"u}ller}, {Florinski}, {Heerikhuisen},
  {Izmodenov}, {Scherer}, {Alexashov}, \& {Fahr}}]{Mueller-etal-2008}
{M{\"u}ller}, H.-R., {Florinski}, V., {Heerikhuisen}, J., {et~al.} 2008, \aap,
  491, 43

\bibitem[{{Opher} \& {Drake}(2013)}]{Opher-Drake-2013}
{Opher}, M., \& {Drake}, J.~F. 2013, \apjl, 778, L26

\bibitem[{{Opher} {et~al.}(2007){Opher}, {Stone}, \&
  {Gombosi}}]{Opher-etal-2007}
{Opher}, M., {Stone}, E.~C., \& {Gombosi}, T.~I. 2007, Science, 316, 875

\bibitem[{{Parker}(1961)}]{Parker-1961}
{Parker}, E.~N. 1961, \apj, 134, 20

\bibitem[{{Pogorelov} {et~al.}(2009){Pogorelov}, {Borovikov}, {Zank}, \&
  {Ogino}}]{Pogorelov-etal-2009}
{Pogorelov}, N.~V., {Borovikov}, S.~N., {Zank}, G.~P., \& {Ogino}, T. 2009,
  \apj, 696, 1478

\bibitem[{{Ratkiewicz} \& {Grygorczuk}(2008)}]{Ratkiewicz-Grygorczuk-2008}
{Ratkiewicz}, R., \& {Grygorczuk}, J. 2008, \grl, 35, 23105

\bibitem[{{Scherer} \& {Fichtner}(2014)}]{Scherer-Fichtner-2014}
{Scherer}, K., \& {Fichtner}, H. 2014, \apj, 782, 25

\bibitem[{{Scherer} {et~al.}(2011){Scherer}, {Fichtner}, {Strauss}, {Ferreira},
  {Potgieter}, \& {Fahr}}]{Scherer-etal-2011}
{Scherer}, K., {Fichtner}, H., {Strauss}, R.~D., {et~al.} 2011, \apj, 735, 128

\bibitem[{Schwadron {et~al.}(2014)Schwadron, Adams, Christian, Desiati, Frisch,
  Funsten, Jokipii, McComas, Moebius, \& Zank}]{Schwadron-etal-2014}
Schwadron, N., Adams, F., Christian, E., {et~al.} 2014, Science, 343, 988

\bibitem[{{Schwadron} \& {McComas}(2013)}]{Schwadron-McComas-2013}
{Schwadron}, N.~A., \& {McComas}, D.~J. 2013, \apj, 764, 92

\bibitem[{{Strauss} {et~al.}(2013){Strauss}, {Potgieter}, {Ferreira},
  {Fichtner}, \& {Scherer}}]{Strauss-etal-2013}
{Strauss}, R.~D., {Potgieter}, M.~S., {Ferreira}, S.~E.~S., {Fichtner}, H., \&
  {Scherer}, K. 2013, \apjl, 765, L18

\bibitem[{{Whang}(2010)}]{Whang-2010}
{Whang}, Y.~C. 2010, \apj, 710, 936

\bibitem[{{Wiengarten} {et~al.}(2014){Wiengarten}, {Kleimann}, {Fichtner},
  {K{\"u}hl}, {Kopp}, {Heber}, \& {Kissmann}}]{Wiengarten-etal-2014}
{Wiengarten}, T., {Kleimann}, J., {Fichtner}, H., {et~al.} 2014, \apj, 788, 80

\bibitem[{{Zank} {et~al.}(2013){Zank}, {Heerikhuisen}, {Wood}, {Pogorelov},
  {Zirnstein}, \& {McComas}}]{Zank-etal-2013}
{Zank}, G.~P., {Heerikhuisen}, J., {Wood}, B.~E., {et~al.} 2013, \apj, 763, 20

\bibitem[{{Zank} {et~al.}(1996){Zank}, {Pauls}, {Williams}, \&
  {Hall}}]{Zank-etal-1996b}
{Zank}, G.~P., {Pauls}, H.~L., {Williams}, L.~L., \& {Hall}, D.~T. 1996, \jgr,
  101, 21639

\bibitem[{{Zhang} {et~al.}(2014){Zhang}, {Zuo}, \&
  {Pogorelov}}]{Zhang-etal-2014}
{Zhang}, M., {Zuo}, P., \& {Pogorelov}, N. 2014, \apj, 790, 5

\end{thebibliography}
\end{document}